\documentclass[12pt,letterpaper]{article}

\usepackage{fixltx2e}
\usepackage{textcomp}
\usepackage{fullpage}
\usepackage{amsfonts}
\usepackage{verbatim}
\usepackage[english]{babel}
\usepackage{pifont}
\usepackage{color}
\usepackage{setspace}
\usepackage{lscape}
\usepackage{indentfirst}
\usepackage[normalem]{ulem}
\usepackage{booktabs}
\usepackage{natbib}
\usepackage{float}
\usepackage{latexsym}
\usepackage{url}
\usepackage{hyperref}
\usepackage{epsfig}
\usepackage{graphicx}
\usepackage{amssymb}
\usepackage{amsmath}
\usepackage{bm}
\usepackage{array}
\usepackage[version=3]{mhchem}
\usepackage{ifthen}
\usepackage{caption}
\usepackage{hyperref}
\usepackage{amsthm}
\usepackage{amstext}

\theoremstyle{plain}

\theoremstyle{definition}

\newcommand\RR{\mathbb{R}}

\newcommand\cT{\mathcal{T}}

\raggedright
\renewcommand{\section}[1]{%
\bigskip
\begin{center}
\begin{Large}
\normalfont\scshape #1
\medskip
\end{Large}
\end{center}}

\renewcommand{\subsection}[1]{%
\bigskip
\begin{center}
\begin{large}
\normalfont\itshape #1
\end{large}
\end{center}}

\renewcommand{\subsubsection}[1]{%
\vspace{2ex}
\noindent
\textit{#1.}---}

\renewcommand{\tableofcontents}{}

\bibpunct{(}{)}{;}{a}{}{,}  

\begin{document}

\bigskip
\medskip
\begin{center}

\noindent{\Large \bf Mean and Variance of Phylogenetic Trees}
\bigskip



\noindent {\normalsize \sc Daniel G. Brown$^1$ and Megan Owen$^2$}\\
\noindent {\small \it 
$^1$David R. Cheriton School of Computer Science, University of Waterloo, Waterloo, ON, N2L 3G1, Canada;\\
$^2$Department of Mathematics, Lehman College - City University of New York, New York, NY, 10468, USA}\\
\end{center}
\medskip
\noindent{\bf Corresponding author:} Megan Owen, Department of Mathematics, Lehman College, City University of New York, 250 Bedford Park Blvd West,
Bronx, NY, 10468, USA; E-mail: megan.owen@lehman.cuny.edu.\\



\subsubsection{Abstract} 
We describe the use of the Fr\'echet mean and variance in the Billera-Holmes-Vogtmann (BHV) treespace to summarize and explore the diversity of a set of phylogenetic trees.  We show that the Fr\'echet mean is comparable to other summary methods, and, despite its stickiness property, is more likely to be binary than the majority-rules consensus tree. We show that the Fr\'echet variance is faster and more precise than commonly used variance measures.  The Fr\'echet mean and variance are more theoretically justified, and more robust, than previous estimates of this type, and can be estimated reasonably efficiently, providing a foundation for building more advanced statistical methods and leading to applications such as mean hypothesis testing. \\
\noindent (Keywords: phylogenetics, treespace, Fr\'echet mean, Fr\'echet variance, two-sample test )\\

\bigskip

Sets of related phylogenetic trees are commonly encountered in evolutionary biology.  For example, one might encounter such a set as the output of an inference program like MrBayes \citep{mrbayes}, or as a set of gene trees on some given set of species.  For this paper, we think of any such set as being a sample from an underlying distribution on the set of all phylogenetic trees with a fixed leaf set, where a tree consists of both a topology and branch lengths.  Here, we consider a mathematically-founded basis for describing the mean of such a distribution, using the representation of trees as elements of a continuous, geometric space and looking for the Fr\'echet mean: the tree that minimizes the sum of the squared distance between the mean and the elements of a sample from the distribution.  This formulation also allows the identification of the Fr\'echet variance, which is the actual sum of the squared distances between the sample elements and the Fr\'echet mean.  

This mean tree has been used in previous work, primarily as a step in the proposal of a more sophisticated statistical approach to analyzing phylogenetic trees \citep{willisConfidence, zairis2016genomic, NyeYoshidaEtal, NyePCA2}.  The variance, or a heuristic computation of it, has been used to compare levels of incongruence under different evolutionary models \citep{williams2012congruent} for hypothesis testing and to validate the application of data cloning to Bayesian phylogenetic inference \citep{PoncianoBurleighEtal}.  However, despite this usage in the literature, there has been no large-scale analysis of the perfomance and properties of the Fr\'echet mean and variance.  

Throughout the paper, we use an iterative algorithm to compute approximations of the Fr\'echet mean and variance, as there is no known polynomial algorithm for computing the mean. This algorithm is fairly efficient, and we show that the mean is close to other established summary measures like the maximum likelihood (ML) tree, maximum a posteriori (MAP) tree, and majority-rules consensus tree.  The mean tree is known to be ``sticky'', a phenomenon in which perturbing a sample may not perturb its mean, and which is caused by negative curvature in the treespace.  This negative curvature can also lead to the mean tree being unresolved, although we show the mean is more resolved than the majority-rules consensus tree, and is predicted to shorten the branch lengths of the mean.  We will refer to the overall property of the mean being pulled towards the star tree as stickiness, and show that even in a best case biological scenario, its effect is still detectable, if minimal, on the mean. 

The Fr\'echet variance of a set of trees quantifies how spread out a set of trees is from their mean. We will show that in our experiments, as sequence length increases and there is more information about the tree to be reconstructed, the variance of samples of trees from the bootstrap and posterior distributions decreases.  We also show that the Fr\'echet variance is a significantly better measure of statistical uncertainty than simpler measures (like the number of topologies found in a set of trees), and is faster to compute than the sum of all pairwise distances between trees, for a large set of trees. Finally we show that the variance can depend on the length of trees in the sample, so this needs to be accounted for in any comparisons of variance.

Our results show that the treespace-based measure of phylogenetic distance that originated with \citet{BHV01} can in fact be used in practical applications. The findings that the Fr\'echet mean and variance make sense biologically, as well as statistically, justify its usage in more sophisticated statistical schemes, such as computing confidence sets \citep{willisConfidence}, and encourage future applications, such as outlier detection.  

\bigskip

\section{Background}
We begin by describing the treespace in which we are working, and some properties of that space and its distance measure.  We also consider other ways of summarizing a collection of trees besides the Fr\'echet mean and variance, and some of their properties.

\subsection{Treespace}
The \emph{Billera-Holmes-Vogtmann (BHV) treespace}, $\cT_n$, \citep{BHV01} contains all unrooted phylogenetic trees with edge lengths and a given set of $n+1$ labelled leaves.  For this paper, we fix the set of leaf labels to be $\{0, 1, ..., n\}$.  Any of these trees can be thought of as rooted by fixing leaf 0 as the root. In this paper, we will define the BHV treespace as a subspace of $\RR^N$, where $N = 2^n - 1$ 
is the number of possible \emph{splits} on $n+1$ leaves, or equivalently, the number of possible partitions of the set of leaves into two sets, each of size at least 1. Each coordinate of $\RR^N$ corresponds to a different split, where the order of the splits does not matter, but is fixed.  Note that by allowing partitions of size 1, we include splits corresponding to edges ending in leaves.  The original definition \citep{BHV01} ignores these edges, but notes that they can be included, as we have done here. 

Given a tree $T$ with $n + 1$ leaves and edge lengths, it corresponds to the following vector in $\mathbb{R}^N$:  for every edge $e$ in $T$ with length $|e|_T$, let the coordinate corresponding to the split induced by $e$ be $|e|_T$.  Let the coordinates corresponding to splits not induced by edges in $T$ be 0.  Let $\cT_n$ be the set of vectors in $\mathbb{R}^N$ that correspond to trees, as just described.  Not all non-negative vectors in $\mathbb{R}^N$ correspond to trees due to split \emph{incompatibility}.  Two splits are \emph{incompatible} if they cannot be induced by edges existing in the same tree.  For example, a \emph{cherry} is a pair of adjacent leaves in a tree, and the corresponding split separates the two adjacent leaves from all others.  No tree with $n \ge 4$ can have both $\{1,2\}$ and $\{1,3\}$ as cherries, so the corresponding splits $\{1,2\}\{0,3, 4, ..., n\}$ and $\{1,3\}\{0,2, 4, 5 ..., n\}$ are incompatible, and no vectors in $\cT_n$ have positive values in both these coordinates.  
The \emph{topology} of a tree is the set of all splits induced by the edges of that tree.  A binary tree, in which all interior nodes have degree 3, contains $2n-1$ splits, while unresolved (or degenerate or non-binary) trees will contain fewer than $2n-1$ splits. 

\begin{figure}[htb]
	\centering
	\includegraphics[scale=0.3]{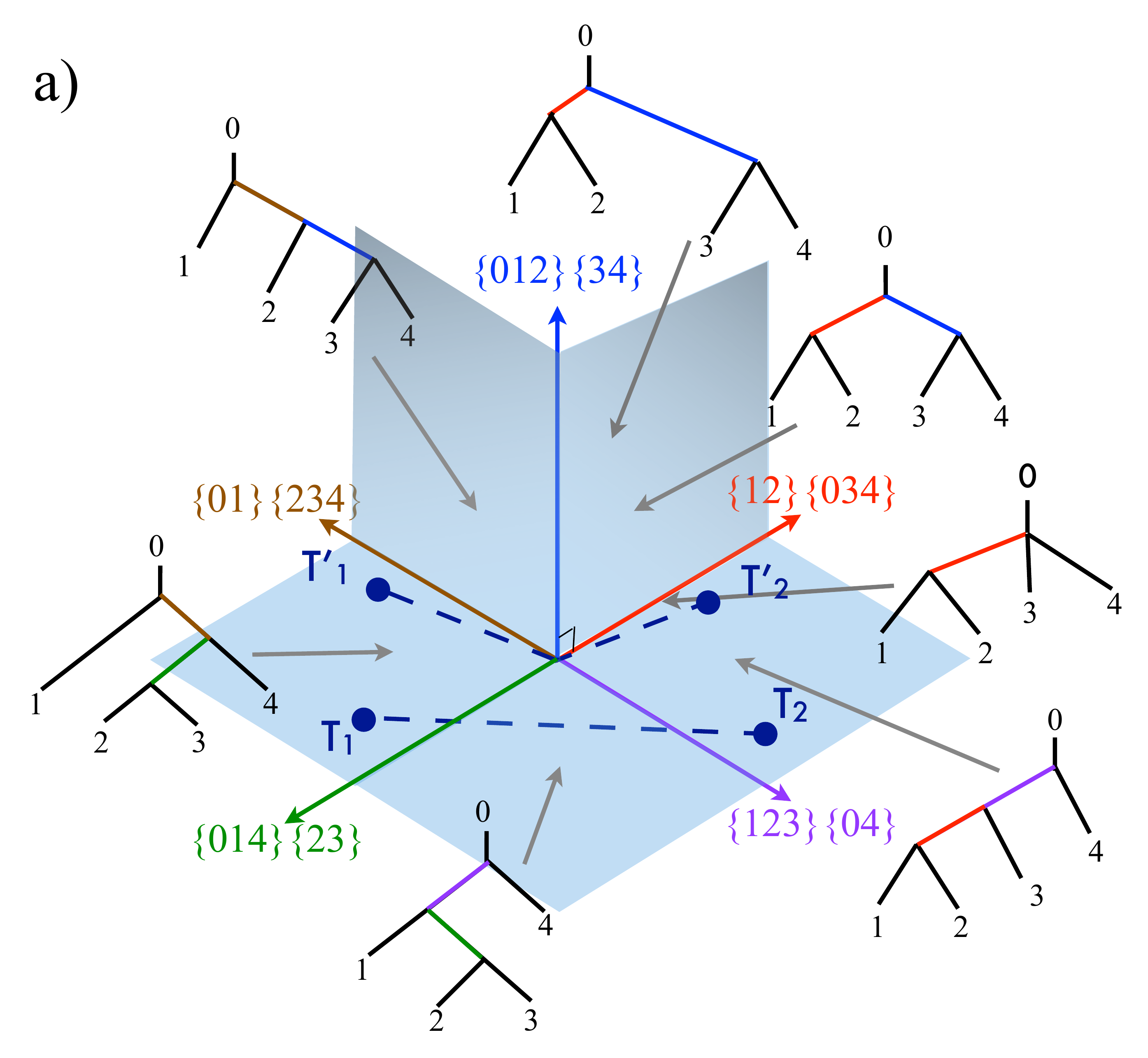}   
	\includegraphics[scale=0.3]{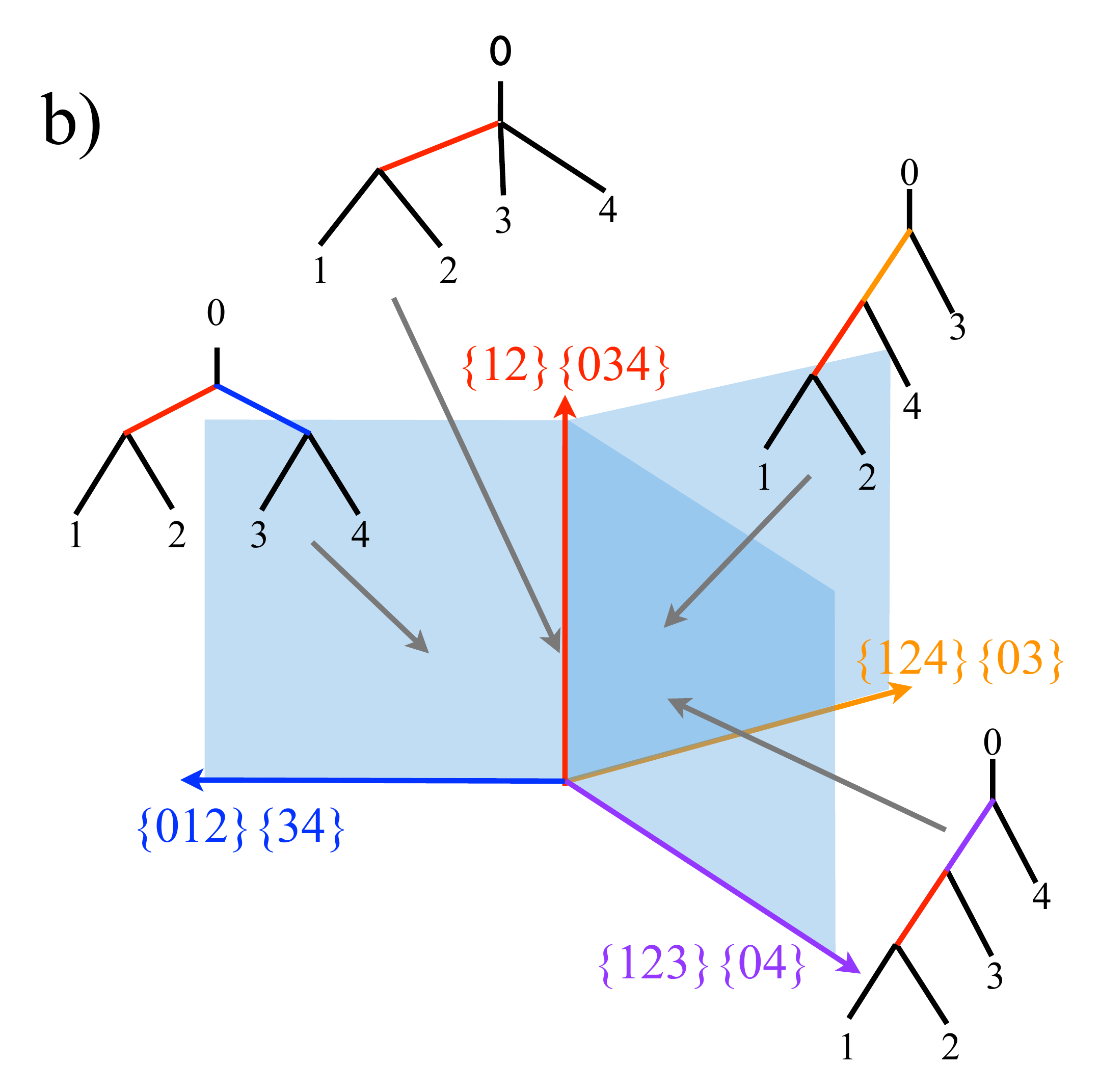}
    \caption{(a) Five and (b) three of the fifteen quadrants in the Billera-Holmes-Vogtmann (BHV) treespace $\cT_4$, with each subfigure representing a different non-Euclidean feature of the BHV treespace.  For ease of visualization, the five dimensions corresponding to the leaf edges are not included. Thus, each binary tree topology is represented by a quadrant (2-dimensional orthant), with the two quadrant axes corresponding to the lengths of the interior edge splits. The axis between two quadrants corresponds to an unresolved tree topology.  The geodesic (shortest path) between two trees is shown by a dashed line and may pass through different orthants depending on the branch lengths of the endpoint trees, as shown by the geodesics between $T_1$ and $T_2$ and between $T_1'$ and $T_2'$.}  
    	\label{fig:5orthants}
\end{figure}

To visualize BHV treespace, consider all trees in $\cT_n$ with the same topology.  Because these trees all correspond to the same set of splits, their vectors have exactly the same set of non-zero coordinates, which can take on any positive values.  Thus, if the number of non-zero coordinates is $d$, this set of trees corresponds to a $d$-dimensional Euclidean \emph{orthant}, which is the non-negative part of $\RR^d$.   When the tree topology is binary, $d = 2n-1$.
There are $(2n-3)!! = (2n-3)\times (2n-5)\times (2n-7) \times \cdot \cdot \cdot \times 1$ binary tree topologies on $n+1$ leaves \citep{Schroder1870}, and thus $(2n-3)!!$ top-level orthants.  Two top-level orthants share a boundary of dimension one less if and only if their corresponding topologies differ by a single Nearest Neighbour Interchange (NNI) move. See Figure~\ref{fig:5orthants}. For more details on the combinatorics and geometry of BHV treespace, see \citet{BHV01}.

There is a distance metric associated with the BHV treespace called the \emph{BHV distance} or the \emph{geodesic distance}, which was also defined in \citet{BHV01}.  For any two trees with the same topology, the BHV distance is the Euclidean distance between their corresponding vectors in their shared orthant.  For two trees with different topologies, the BHV distance is the length of the shortest path between them that remains in the treespace. The length of any path can be computed by calculating the Euclidean distance of the path restricted to each orthant that it passes though, and summing these lengths. The shortest path is called a \emph{geodesic}, and will pass from one orthant to the next orthant through lower-dimensional boundaries corresponding to trees with fewer splits.  See Figure~\ref{fig:5orthants} for an example of two geodesics in $\cT_4$.

The BHV treespace is connected, because there is a path between any two trees that consists of the line segment from the first tree to the origin, followed by the line segment from the origin to the second tree.  This path through the origin may or may not be the geodesic.  \citet{BHV01} showed that the BHV treespace is globally non-positively curved \citep{BH99}, which implies that geodesics are unique.  \citet{OP11} gave a polynomial time algorithm for computing the geodesic distance between two trees that runs in $O(nm^3)$ time, where $m$ is the number of leaves in the largest subtree formed by decomposing the trees along common splits.

\subsection{Mean and Variance}
In Euclidean space, the Fr\'echet mean, or centre of mass, is the point minimizing the sum of the squared distances to the sample points, and is equivalent to the coordinate-wise average of the sample points.  The Fr\'echet mean was similarly defined for treespace by \citet{MillerOwenProvan2015} and \citet{Bacakmeanmedian}, independently.  For a set of sample trees $\{T_1, T_2, ..., T_r\}$, the Fr\'echet mean, or simply, \emph{mean}, is the tree $t$ which minimizes $\sum_{i=1}^r d(t,T_i)^2,$ where $d(\cdot,\cdot)$ is the BHV distance between the two trees.  The Fr\'echet variance, or simply \emph{variance}, is that minimum sum of squared distances.  This mean is unique because treespace is non-positively curved.  Both \citet{MillerOwenProvan2015} and \citet{Bacakmeanmedian} gave an iterative algorithm for approximating the mean and variance based on the Law of Large Numbers for non-positively curved space derived by \citet{Sturm03}.  

\begin{figure}[htb]
	\centering
	\begin{tabular}{ccc}
	\includegraphics[scale=0.4]{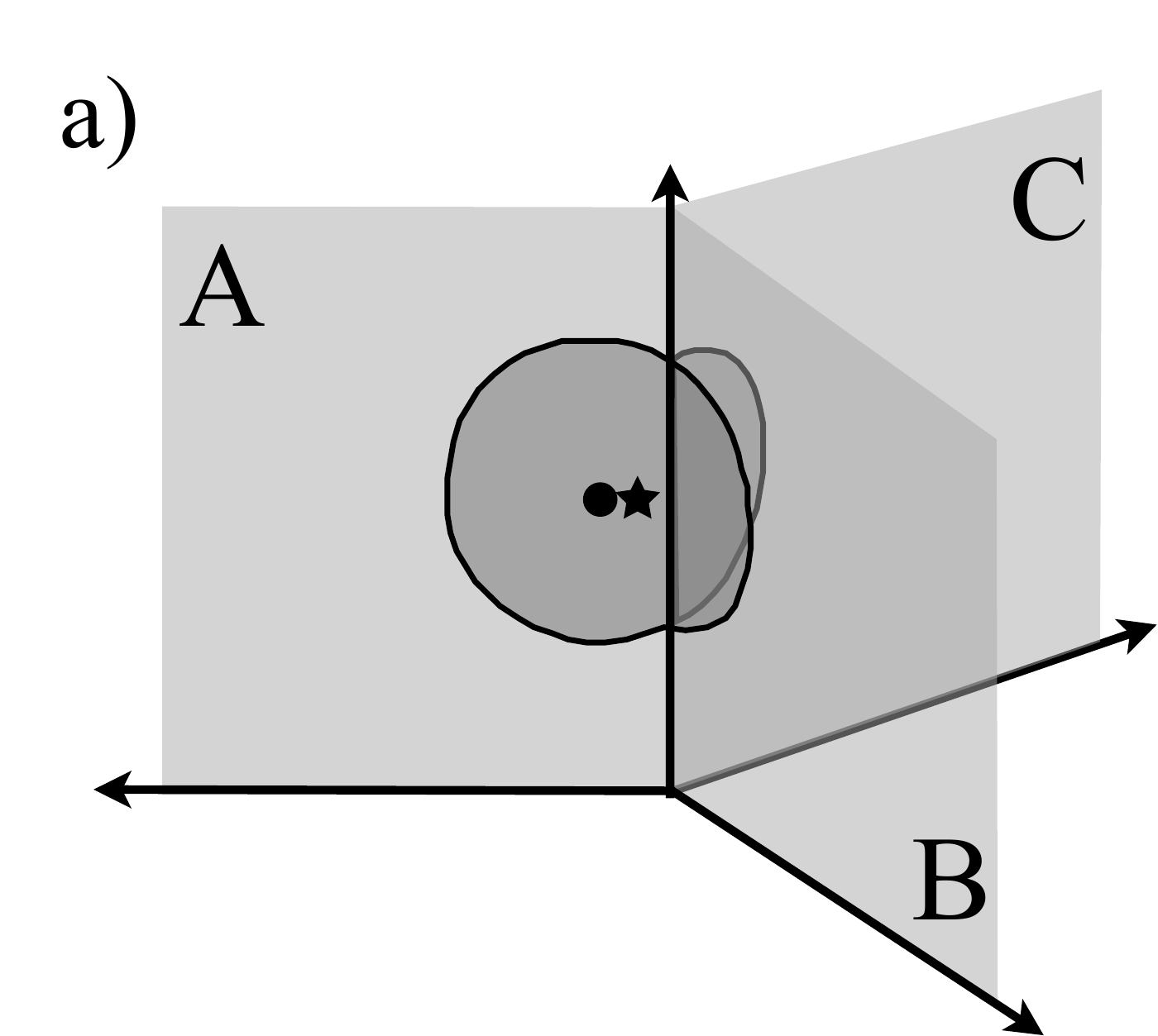}&&   
	\includegraphics[scale=0.4]{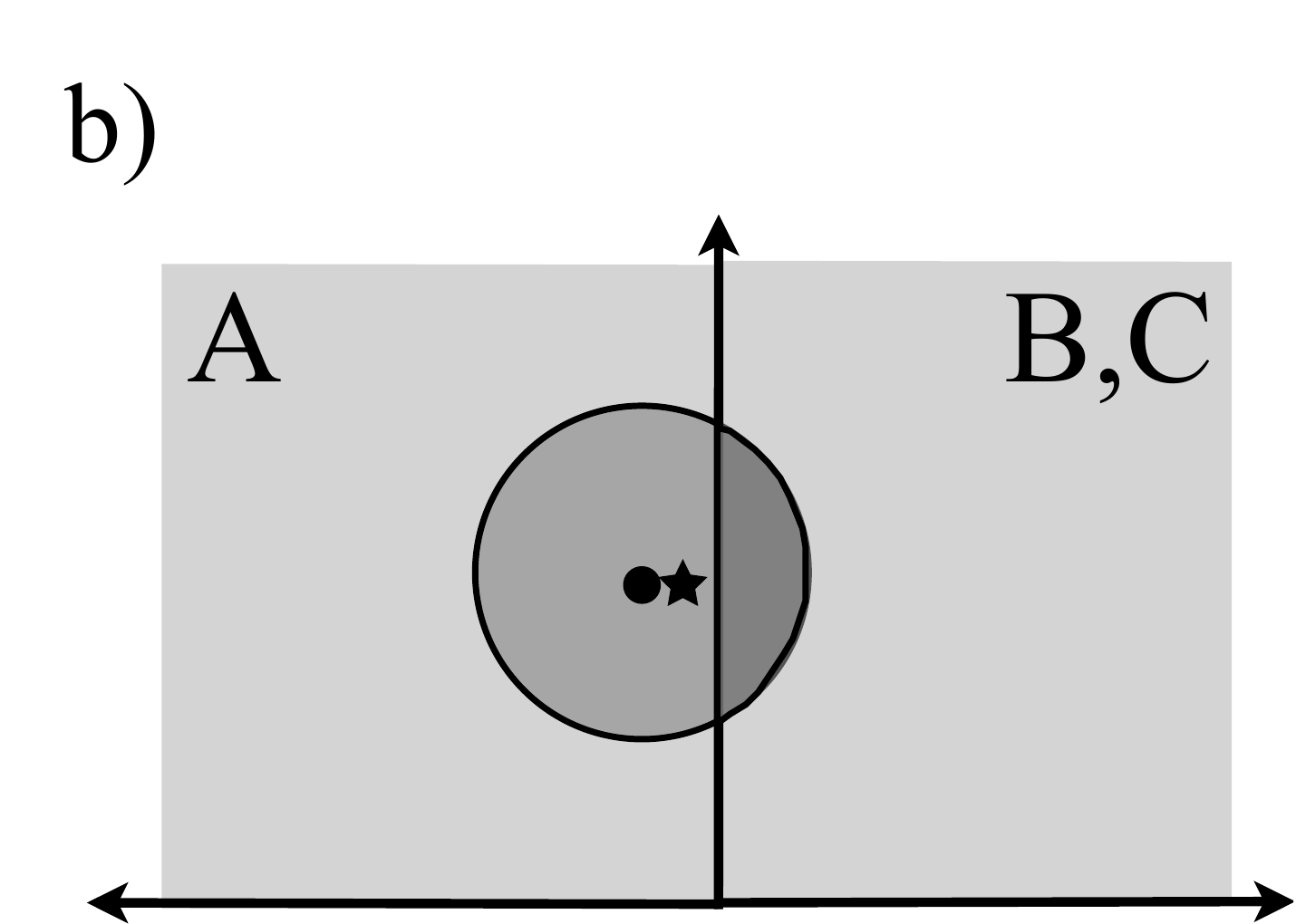}
	\end{tabular}
    \caption{(a) Three neighbouring quadrants in BHV treespace, labelled $A$, $B$, and $C$.  We uniformly sample from a ball of radius 1 around the centre tree (marked by a dot) in quadrant $A$, as shown in (a).  Flattening quadrants $B$ and $C$ gives (b), which shows that computing the mean of this distribution (marked by a star) is equivalent to computing the weighted Euclidean mean of a disk with twice the density in the region where the radius 1 ball intersects orthants $B$ and $C$.}  
    	\label{fig:sticky}
\end{figure}

We briefly mention some interesting properties of the mean.  See \citet[Section 5]{MillerOwenProvan2015} for details and proofs. The mean tree is not necessarily a refinement of the majority-rule consensus tree (which contains all splits found in a majority of the trees $T_i$). Any split that appears in the mean tree appears in at least one of the sample trees, and any split that appears in all the sample trees appears in the mean tree.  Finally, the mean tree is ``sticky'' \citep{hotz2013sticky}, in that perturbing one of the sample trees does not always change the mean tree. This ``stickiness'' only happens when the mean is on a lower-dimensional orthant in treespace, which corresponds to an unresolved tree topology.  Thus, the mean will be unresolved more often than one might expect or wish, similar to the majority-rule consensus tree.  Stickiness is caused by the non-Euclidean nature of BHV treespace, and also causes binary mean trees to be pulled towards the lower-dimensional orthants.  For example, suppose we wish to find the mean of a set of trees sampled uniformly from a ball of radius 1 around a centre tree, as in Figure~\ref{fig:sticky}a.  By folding the orthants $B$ and $C$ on top of each other, we see that in the limit as the sample size increases, computing this mean is equivalent to computing the weighted Euclidean mean, or centre of gravity, of a disk in which a section corresponding to the part of the ball in orthants $B$ and $C$ has twice the density of the rest of the disk (see Figure~\ref{fig:sticky}b).  The mean will be closer to the shared axis than the original centre tree.  This behaviour is also a component of the stickiness of the mean.

\citet{Sturm03} showed that a Law of Large Numbers holds, meaning that as the sample size increases, the sample means of a distribution over $\cT_n$ converge to the true mean.  \citet{CLTcomplete} proved a Central Limit Theorem on the BHV treespace, showing that the distribution of the sample means converges to a certain Gaussian distribution.  This Central Limit Theorem was used by \citet{willisConfidence} to construct confidence sets in BHV treespace.



\subsection{Other Measures of Centrality and Variance} 

We compare the Fr\'echet mean to three other commonly used measures of centre or consensus in phylogenetics, of which some are only the topology and others are both the topology and edge lengths.  The first is the majority-rule topology, which is the topology containing exactly those splits appearing in a majority of the input trees \citep{majority-rule}. The majority-rule consensus topology and the Fr\'echet mean are not, in general, refinements of each other \citep{MillerOwenProvan2015}.

The other two measures of centre come from tree search algorithms, namely maximum likelihood (ML) and Bayesian inference.  Both tree search procedures produce a distribution of trees along with a most likely or most probable tree, which acts like a centre.  Specifically, maximum likelihood search produces a distribution of bootstrap trees \citep{bootstrap} and a maximum likelihood (ML) tree, while Bayesian inference produces a posterior distribution and a maximum a posteriori (MAP) topology \cite{rannala1996probability}.

The variance or amount of variability in a set of trees is less established as a measure than the summary tree.  Often, a visual representation, such as SplitsTree \citep{splitstree4} or DensiTree \citep{densitree}, is used to represent the diversity of a set of trees.  Unfortunately, these methods cannot be assessed or compared quantitatively.  Instead, we will compare the Fr\'echet variance with several quantitative measures of variance, namely the number of different tree topologies in the set, the number of different splits in the set, and the sum of the squared geodesic distances between each pair of trees in the set.  The later measure was proposed in \citet{CH}. 

\subsection{Relationship between the BHV and Robinson-Foulds Distances}

The BHV distance and weighted version of the Robinson-Foulds (RF) distance \citep{RF} are closely related, leading to relationships between measures of centre involving these distances.  The RF distance between trees $T_1$ and $T_2$ is defined as the number of splits in $T_1$ but not in $T_2$, plus the number of splits in $T_2$ but not in $T_1$.  That is, it is the size of the symmetric difference between the split sets of $T_1$ and $T_2$.  Note that this definition only uses the tree topologies, and not the edge lengths of the trees.  The weighted Robinson-Foulds (WAF) distance \citep{weightedRF} is the sum of the lengths of the splits in only one of the trees, plus the difference in the lengths of the common splits.  If the $L_1$ metric, which is also known as the Manhattan or taxicab metric, is used in each orthant of BHV treespace instead of the $L_2$ or Euclidean metric, then the length of a (no longer unique) geodesic between two trees is the same as the weighted RF distance between them (see \citet{StJohnReview} for a good summary).  Because the WAF and BHV distances only differ by whether the $L_1$ or $L_2$ metric is used in each orthant, pairs of trees that are close under one are close under the other.  Thus, results validating the WAF distance for biological application \citep{kuhner2014practical} are likely to carry over to the BHV distance.  While the BHV distance is more expensive to compute, BHV treespace has mathematical properties, like the uniqueness of geodesics and means, that make it more appropriate for developing further statistical tools for analyzing trees.

Recall that the median tree is the tree minimizing the sum of distances, instead of the sum of squared distances as in the mean tree, to the sample trees.  The majority-rule consensus tree is the median tree under the Robinson-Foulds distance \citep{BarthelemyMcMorris}, and \citet{Pattengale} further showed that the weighted majority-rule consensus tree is the median under the weighted Robinson-Foulds distance.

\subsection{Related Work}
The most closely related work to ours is that of \citet{BennerBacakBourguignon}.  They investigated the behaviour of both the mean and the median (the tree minimizing the sum of distances, instead of the sum of squared distances, to the sample trees) in summarizing posterior distributions returned by Bayesian tree reconstruction methods. The authors simulated sequences of lengths 50, 100, 250, and 500 using a 14-taxa tree of plants and the F81 evolutionary model \citep{F81}.  They show that the mean and median estimates are comparable to the majority-rule consensus estimate, and in some instances perform better, and also investigate how the Fr\'echet variance changes with sequence length.

Our work was conceived independently, and while the overall aim of the work is the same, the scope of our experiments are much broader, considering tree distributions generated from both maximum likelihood and Bayesian methods; the general GTR evolutionary model; trees with more taxa; stickiness of the mean in practice; and a more comprehensive look at the Fr\'echet variance, including its relationship with tree length.

This paper focuses on the Fr\'echet mean and variance of a set of trees, with the mean being a point summary of the data.  Other work has looked at 1-dimensional summaries, or best fit lines, in BHV treespace \citep{NyePCA1,NyePCA2,IPMI}, and extensions to multi-dimensional summaries or a generalization of Principal Components Analysis (PCA) \citep{NyeYoshidaEtal}.

The Fr\'echet mean is not the only mathematically justified summary statistic for a set of trees.  \citet{holder2008justification} showed that the majority-rules consensus tree is the optimal tree to report as a summary of a Bayesian posterior when using a loss function that is linear in the number of incorrect clades estimated and number of true clades missing from the estimate, and that treats an incorrect split as a more serious error than an omitted split.  \citet{huggins2011bayes} continued this line of investigation by showing that using the path difference distance \citep{steel1993distributions} between trees as the loss function in the Bayes Estimator applied to a posterior distribution sample improved the accuracy of the reconstruction tree.  When the loss function in a Bayes Estimator is a distance, computing the Bayes Estimator becomes equivalent to computing a median.

\bigskip
\section{Methods}
\label{sec2}

\subsection{Simulated Data} \label{s:simulation_method}

\begin{figure}[htb]
    \centering
    \includegraphics[scale=0.4]{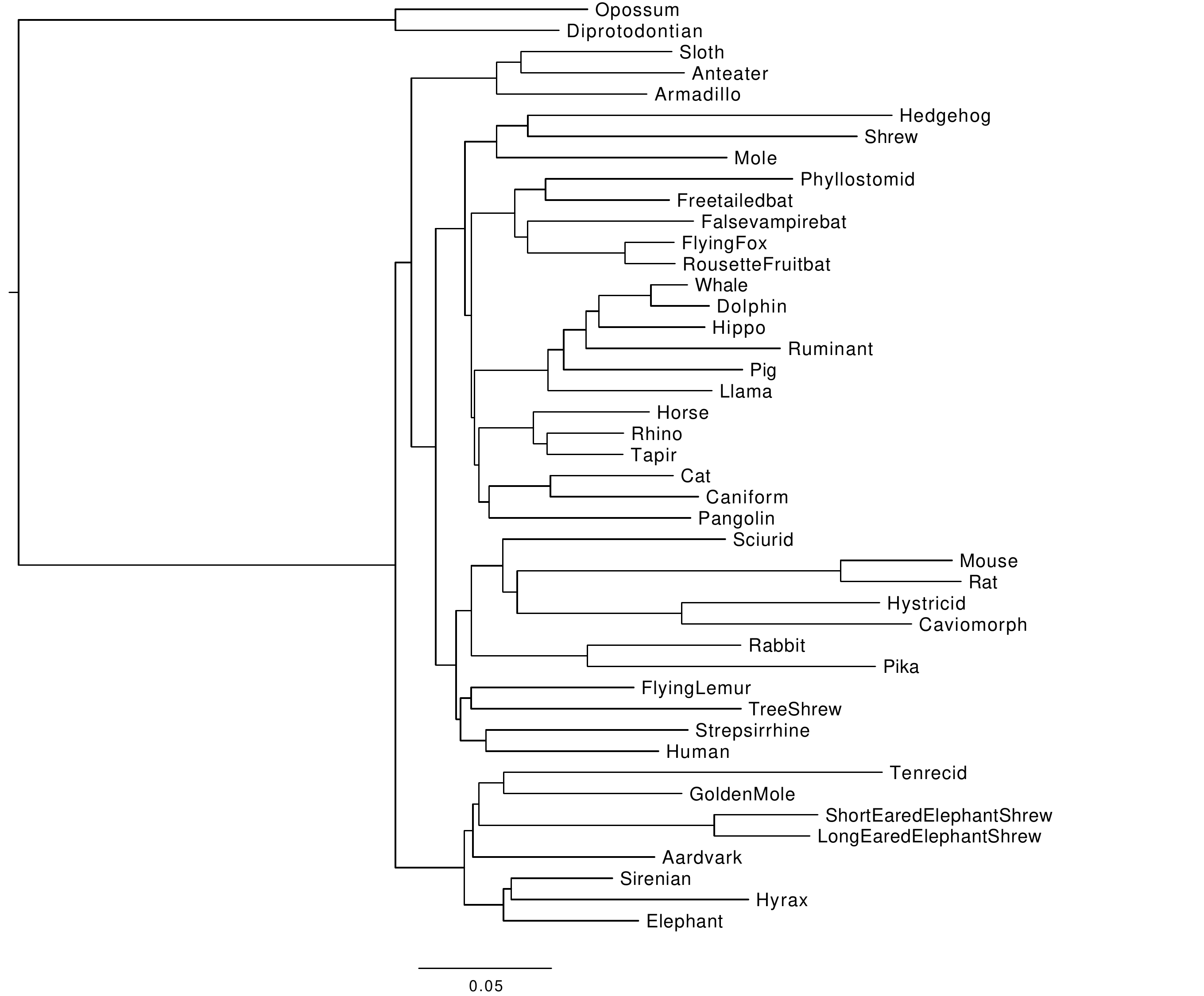}
    \caption{Reference mammal tree from \citet{Murphy01}.}
    \label{f:orig_tree}
\end{figure}

For the first dataset, MAMMAL, we simulated 10 sets of DNA sequences for a variety of lengths (500, 1000, 1500, 2000, 2500, 3000, 3500, 4000 base pairs), using the tree of 44 mammals from \citet{Murphy01}, which is shown in Figure~\ref{f:orig_tree}, as the reference tree.  The sequences were simulated with Seq-Gen version 1.3.3 \citep{seq-gen} using the GTR + I model with the following parameter settings:  the proportion of invariable sites is 0.18; the equilibrium frequencies for A, C, G, T are 0.21, 0.31, 0.3, and 0.18, respectively; and the GTR rate matrix is $(1.5,4.91,1.34,0.83,5.8,1)$, where the entries are the transition rate from A to C, A to G, A to T, C to G, C to T, and G to T.  These parameter values were estimated in \citet{Hillis} for this reference tree.

\begin{figure}[htb]
    \centering
    \includegraphics[scale=0.5]{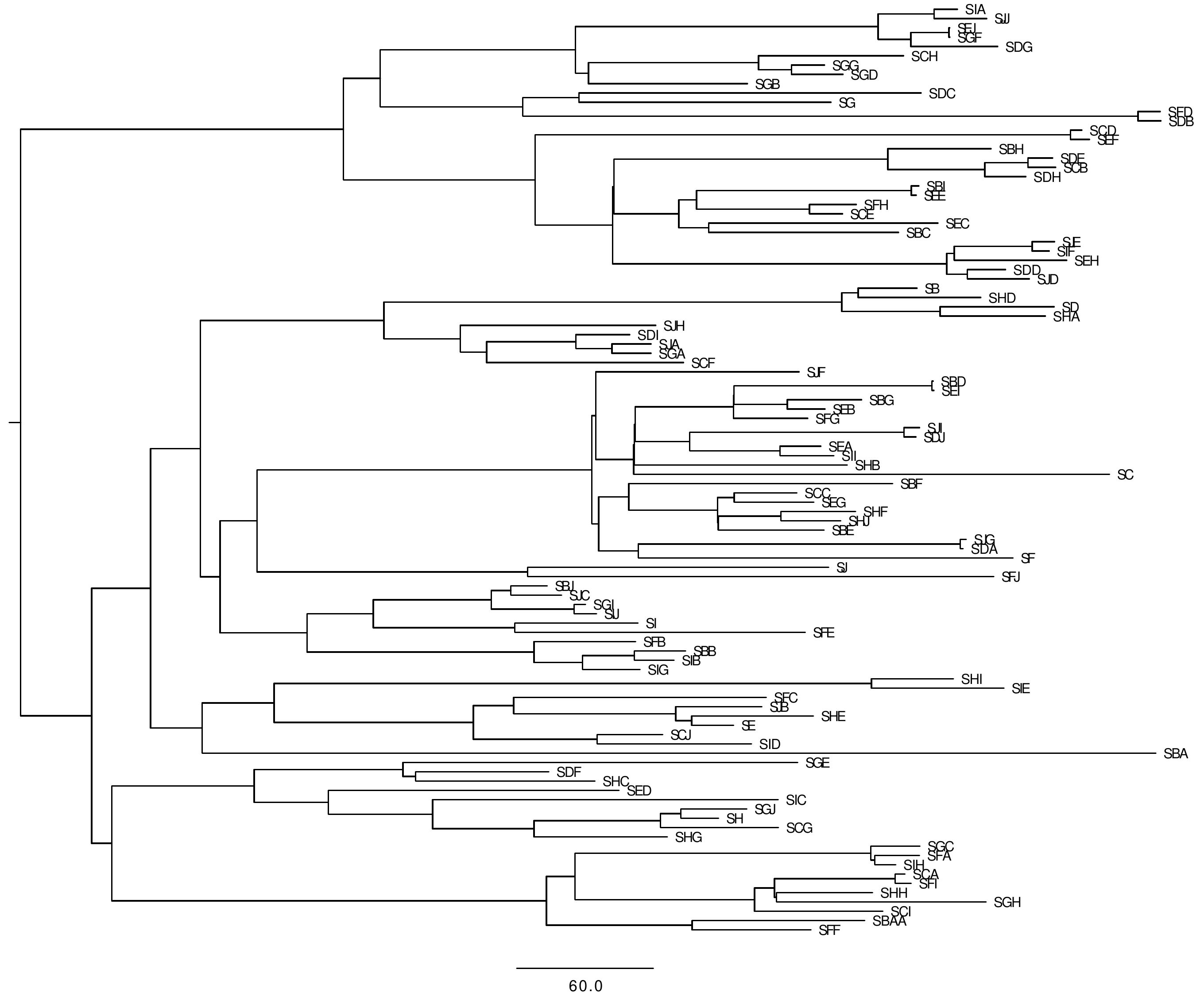}
    \caption{Example SATE reference tree for 100S1 model, first replicate.}
    \label{f:sate_ref_tree}
\end{figure}

The second dataset, SATE, was the sets of 100 taxa sequences simulated for, and used in, \citet{sate}.  These sequences were produced by generating reference trees using r8s \citep{r8s}, which were then adjusted to not be ultrametric and each edge length was scaled by a ``tree height" value.  Sequences of 1000 base pairs were simulated via ROSE \citep{rose} using the GTR + Gamma model and either short, medium, or long gaps.  There were 20 replicates, each with their own reference tree, for the seven different model conditions.  See Figure~\ref{f:sate_ref_tree} for an example of a reference tree, and the supplemental material of \citet{sate} for complete details.

For each set of sequences in each dataset, we ran RAxML Version 8 \citep{raxml_v8} to compute the Maximum Likelihood (ML) tree and 1000 bootstrap sample trees. For the RAxML settings, we used the GTR + I + Gamma evolutionary model.  We conducted a full analysis (option \verb+-f a+) using rapid bootstrapping (option \verb+-x+), which was recommended in the RAxML user manual 
when computing a large number of bootstrap replicates, since every 5th bootstrap tree is used as a starting point for the ML tree search.
For each set of sequences, we also ran MrBayes 3.2 \citep{mrbayes} to compute the MAP tree and sample 1000 trees from the posterior distribution.  For the MrBayes settings, we used the GTR+I+Gamma evolutionary model, and ran for 5,000,000 iterations, sampling every 1000 generations. We used the last 1000 sampled trees as the posterior distribution.  We verified convergence of each run by comparing the log likelihood values the two chains.

For each sample distribution of 1000 trees, we estimated the Fr\'echet mean and variance, using SturmMean \citep{MillerOwenProvan2015} with the settings explained in the following subsection.  We also computed the majority-rule consensus tree using Dendropy \citep{dendropy}, and three other measures of variation for each sample: the number of different tree topologies in the sample, the number of different splits in the sample, and the sum of the squared BHV distance between each pair of trees in the sample.  We compared the mean tree to the corresponding reference, majority-rule consensus, ML, and MAP trees using the RF distance and the BHV distance if both trees had meaningful branch lengths.  The mean trees, ML trees, and reference trees all had meaningful branch lengths, while the majority-rule consensus tree and MAP trees did not.

For MAMMAL, we compared the number of unresolved mean and majority-rules trees for each sequence length, as well as the lengths of the sampled trees and their centres.  Here we define the length of a tree $T$ with edge set $E$ to be $\sqrt{\sum_{e \in E} |e|_T^2}$ which is also the BHV distance of tree $T$ to the BHV treespace origin.  To investigate stickiness of the mean, we compare
the length of the mean tree of a sample with the average length of trees in that sample, to see if the mean is much shorter than its component trees.  We also compare the lengths of the mean and the average lengths of trees between the posterior and bootstrap distributions generated from the same sequence set.  In all cases, we use the Wilcoxon ranked signed test \citep{WilcoxonSignedRankTest} to test the hypothesis that the two distributions of lengths are the same.

Finally, we conducted a mean hypothesis test on the bootstrap and posterior samples for the first repetition of 4000 base pairs in MAMMAL.  Recall that a mean hypothesis test is a type of two-sample test that tests if the means of two samples are the same.  Rejecting this hypothesis implies that the samples are from different distributions.  We computed the BHV distance between the means of these bootstrap and posterior samples, and compared it to the BHV distance between the means of a random partition of the two samples.  This random partition was made by combining the bootstrap and posterior trees into one set, and randomly partitioning this new set into two equal parts.
Since a full permutation test is not feasible (since each sample contains 1000 trees), we estimated it using 500 randomly chosen permutations.  Note we are following the method for performing two-sample hypothesis testing on trees that was suggested in \citet{IPMI} for lung-airway tree data.

\subsection{Computation of the Mean and Variance}
We compute an approximation of the mean tree using the iterative implementation described in \citet{MillerOwenProvan2015}.  In this implementation, a new approximation of the mean tree is returned each iteration, and these approximations converge to the true mean tree as the number of iterations grows.  To decide when to stop the iterative algorithm, we use a program option to check for convergence using a Cauchy sequence of length 10 with an epsilon of $10^{-6}$ for MAMMAL and $10^{-4}$ for SATE.  In other words, we stop the iterative algorithm when the pairwise BHV distances between the last 10 mean approximations were all less than or equal to this epsilon.  In all of our experiments, the means converged within 285 000 iterations. We also use the random permutation heuristic, which selects the tree used at each iteration randomly without replacement instead of with replacement.  This heuristic improves the convergence time of the algorithm in practice.  We validated our choice of epsilon for MAMMAL by choosing one repetition for each sequence length, and computing the (approximate) mean of the corresponding posterior distribution sample 10 times using the chosen parameters.  For an epsilon of $10^{-6}$, the average BHV distance between pairs of approximate means of the same sample was on the order of $10^{-5}$, which we considered acceptable. The nature of the iterative approximation algorithm causes the estimated mean trees to be binary.  We declared an estimated mean tree to be unresolved if one or more of its internal edges were less than epsilon in length. Computing the mean and variance for MAMMAL took 10-35 minutes on a 3.5 GHz 6-Core Intel Xeon E5 Processor.  The program and source code are available at \url{ http://comet.lehman.cuny.edu/owen/code/SturmMean.tar.gz}.

\bigskip
\section{Results}  
\begin{figure}[!htb]
\centering
\includegraphics[scale=0.8]{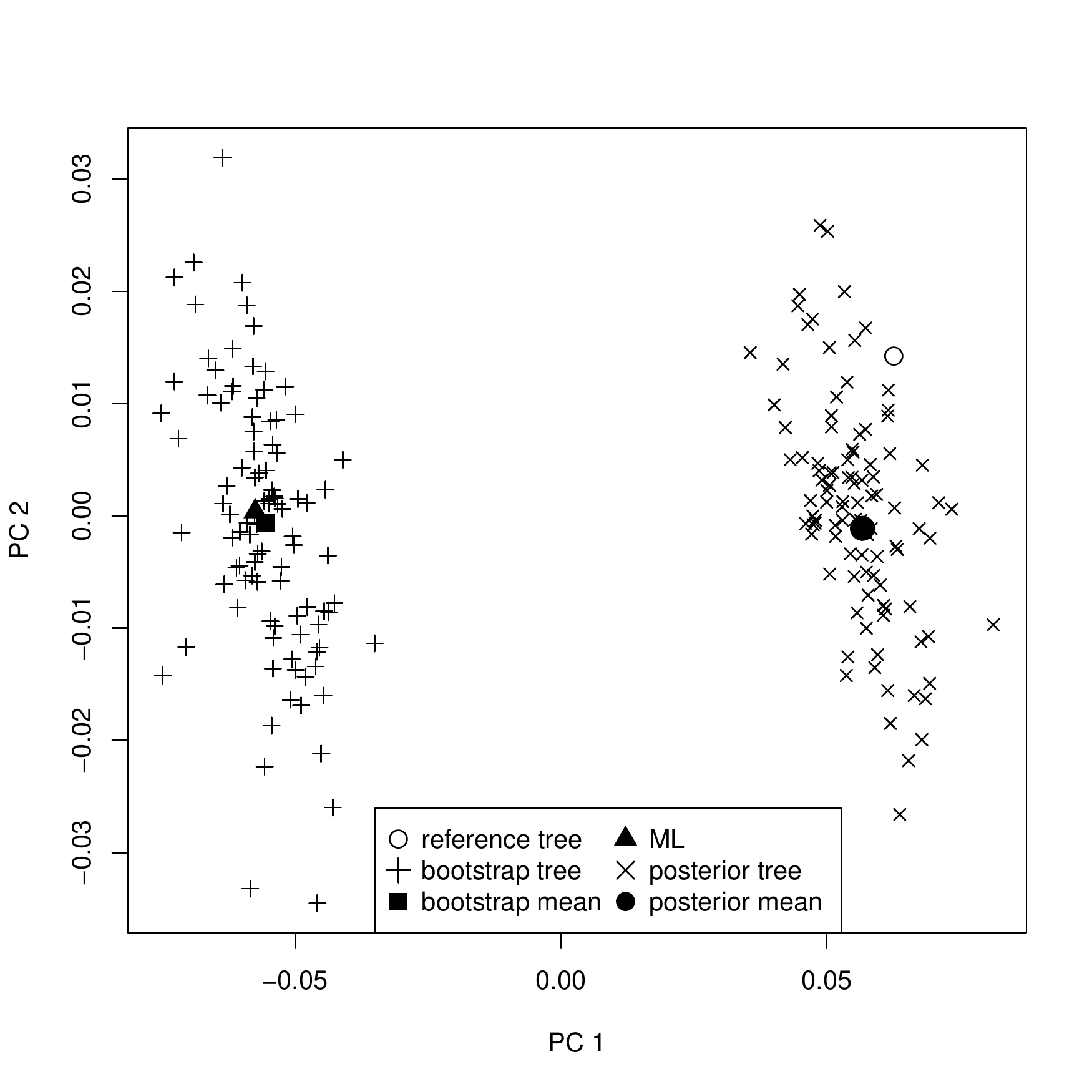}
\caption{A sample of trees from the first repetition of the 4000 base pair sequence length from the MAMMAL dataset embedded in 2 dimensions using classic Multi-dimensional Scaling. The trees from the bootstrap and posterior samples form two clusters, most likely due to differences in branch lengths instead of topology.}
\centering
\label{f:mds}
\end{figure}

To visualize what the trees from one repetition of the MAMMAL dataset look like, for the first repetition of the 4000 base pair sequence length experiment, we computed the BHV distance between 100 trees from each of the bootstrap and posterior samples, the reference tree, the ML tree and the two mean trees.  We use classic Multi-dimensional Scaling (MDS) \citep{kruskal1964mds} to reduce this to two dimensions (Fig.~\ref{f:mds}).  The two means and the ML tree are in the middle of their respective samples.  The two samples are separated in space and the reference tree is closer to the posterior sample, but not near its centre. The reference tree, two means, ML tree, and 62 of the trees from the two samples have the same topology.  The other common topologies appear in both the bootstrap and posterior samples, suggesting that the difference between the two clusters in Figure~\ref{f:mds} is primarily due to branch lengths, which the BHV distance takes into account, instead of topology.

\subsection{Comparison of the mean tree with other measures of centre}
\begin{figure}[!htb]
\centering
\includegraphics[scale=0.4]{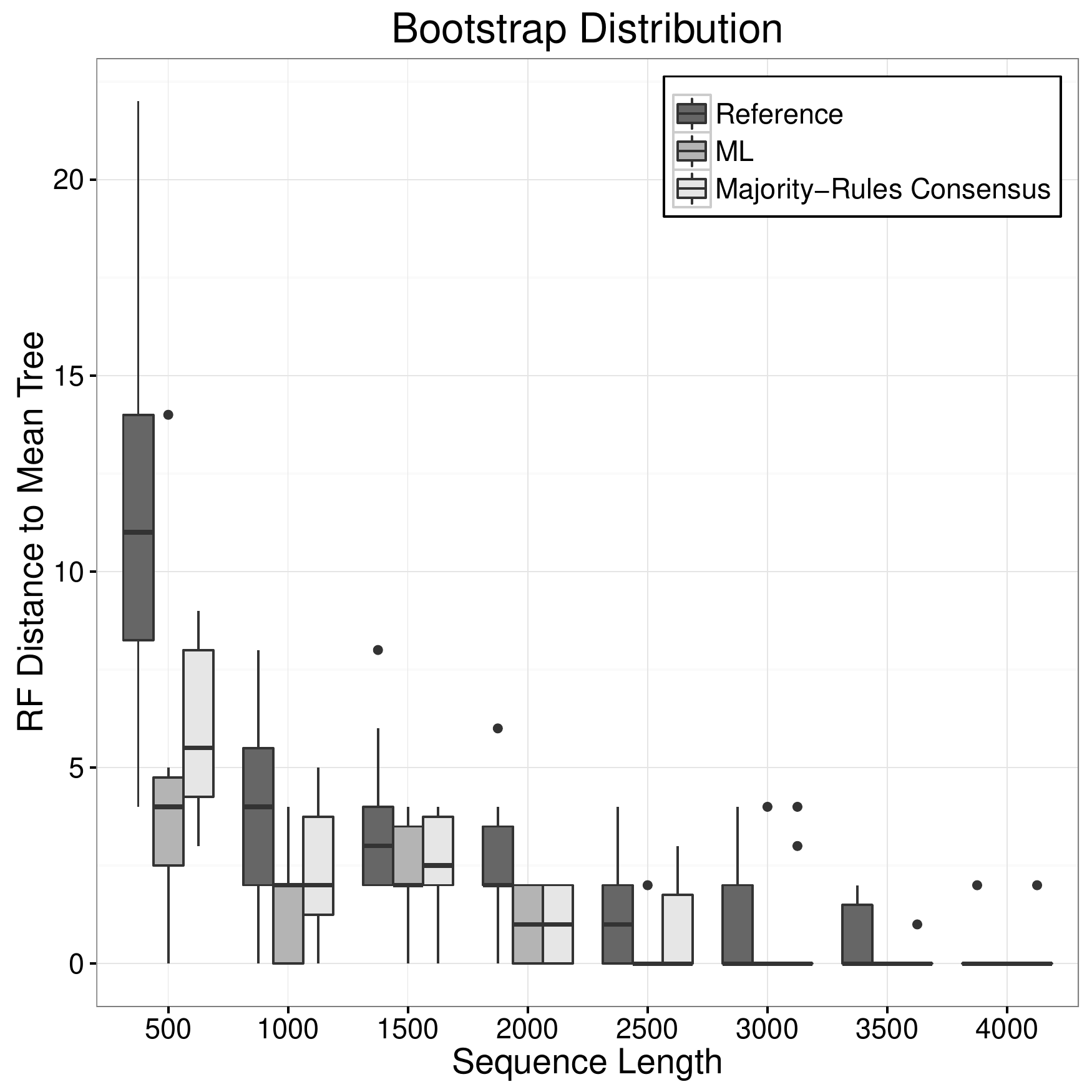}
\includegraphics[scale=0.4]{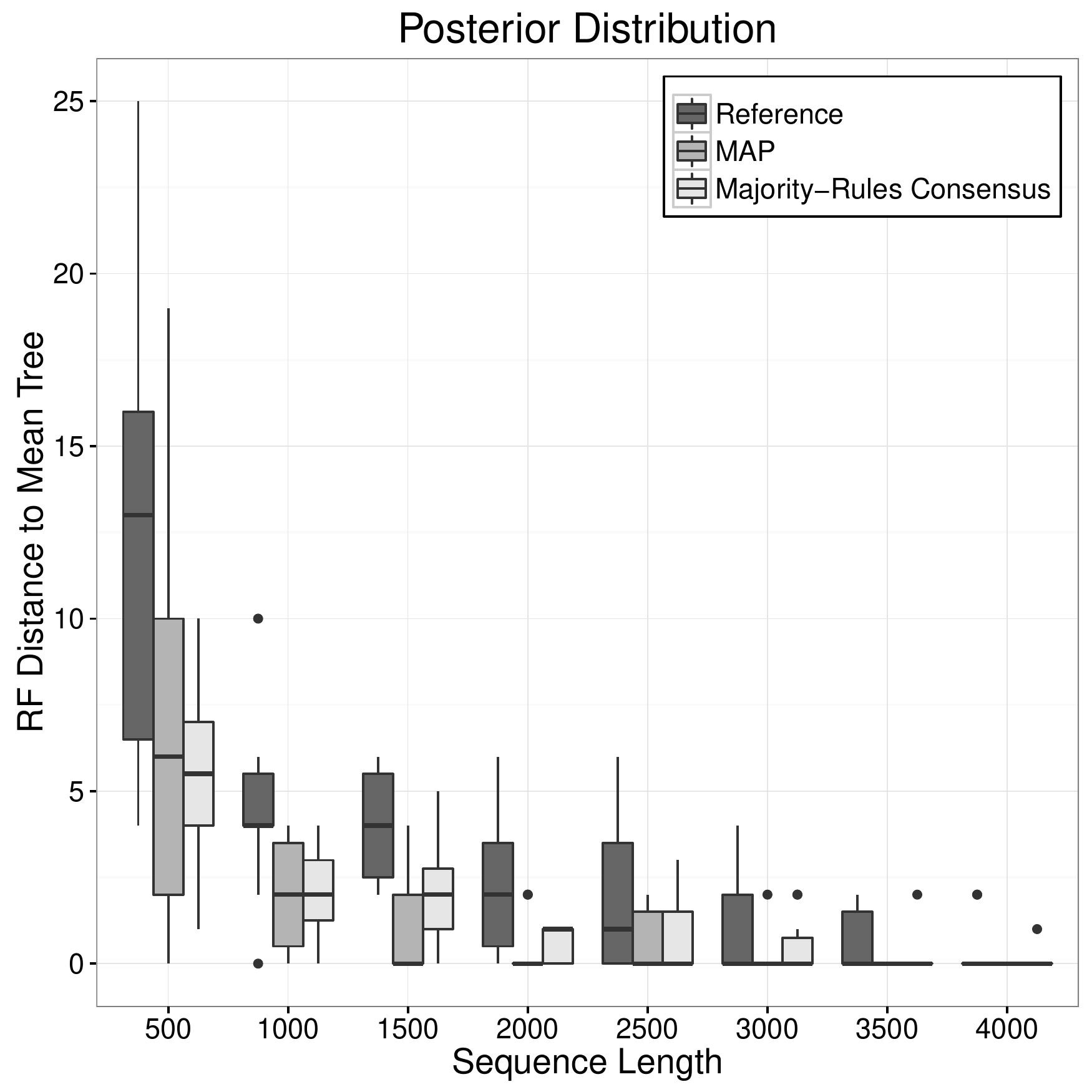} 
\includegraphics[scale=0.4]{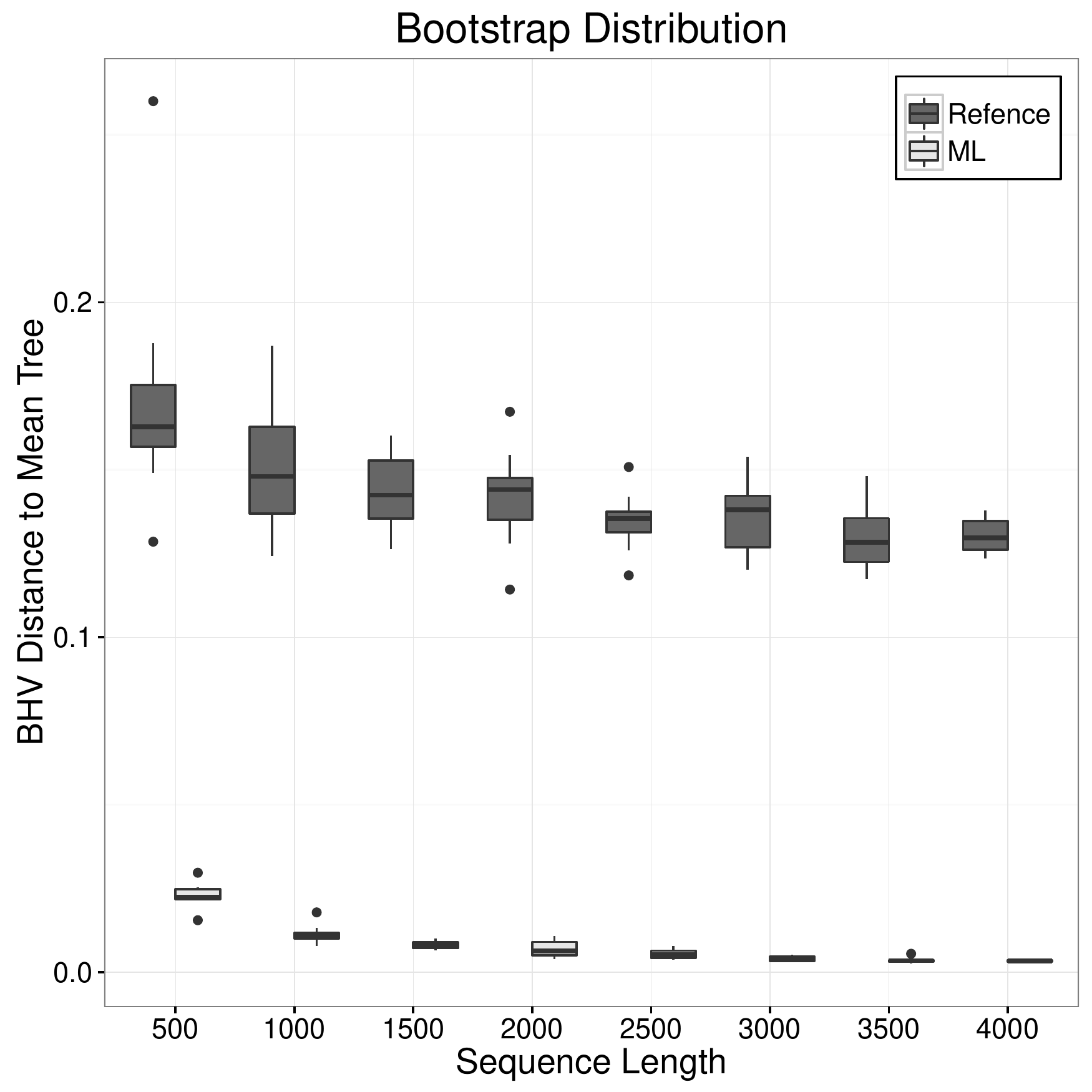} 
\includegraphics[scale=0.4]{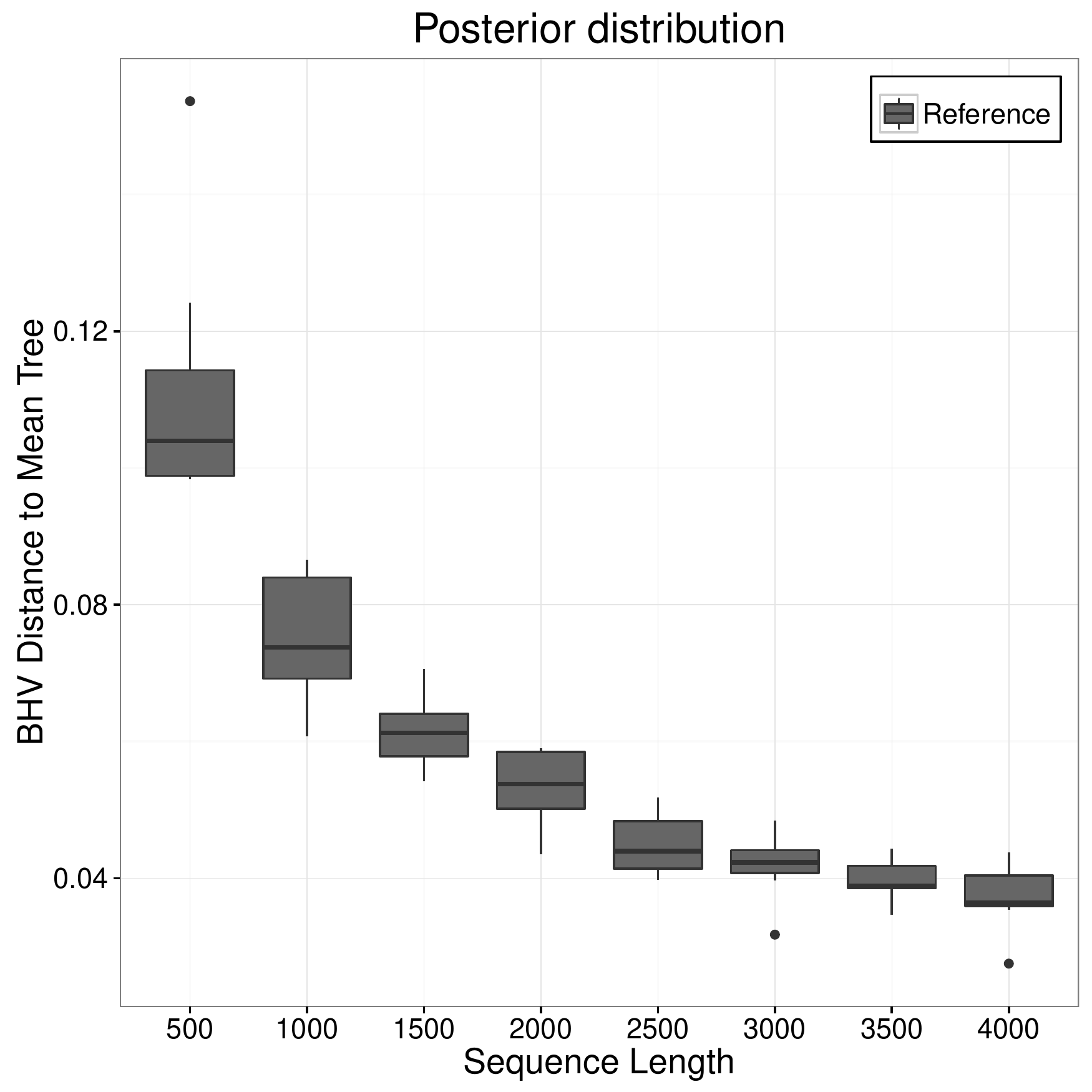} 

\caption{For each sample generated from the MAMMAL dataset, the RF and BHV distances were calculated from its mean tree to its majority-rule tree, ML or MAP tree, and the reference tree. The distribution of these distances is shown using box plots. The mean tree approaches the ML or MAP, and consensus trees as the sequence length increases, with all three trees usually sharing the same topology by sequence lengths of 3000 base pairs.}
\centering
\label{f:mammal_mean_to_all}
\end{figure}

\begin{figure}[!htb]
\centering
\includegraphics[scale=0.4]{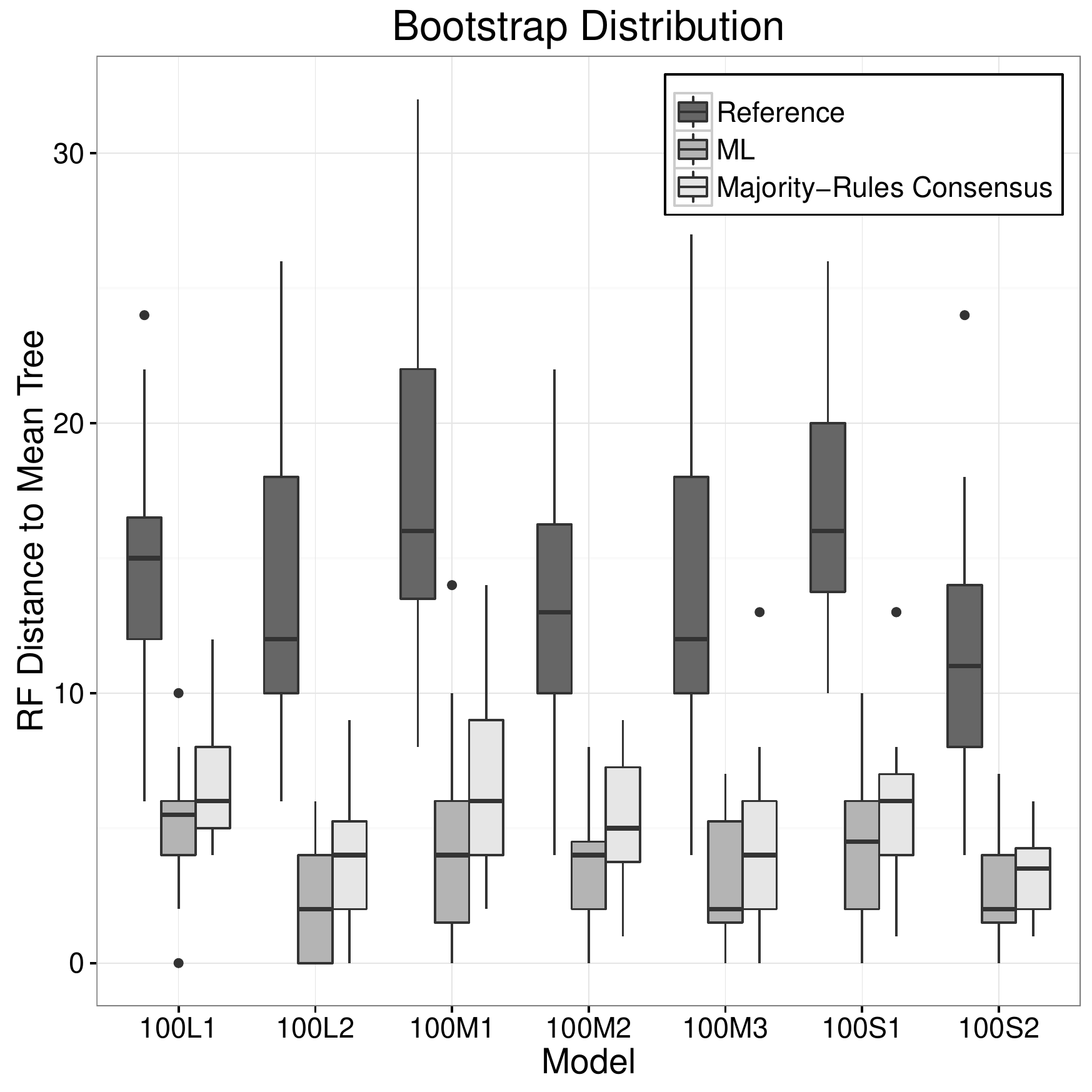}
\includegraphics[scale=0.4]{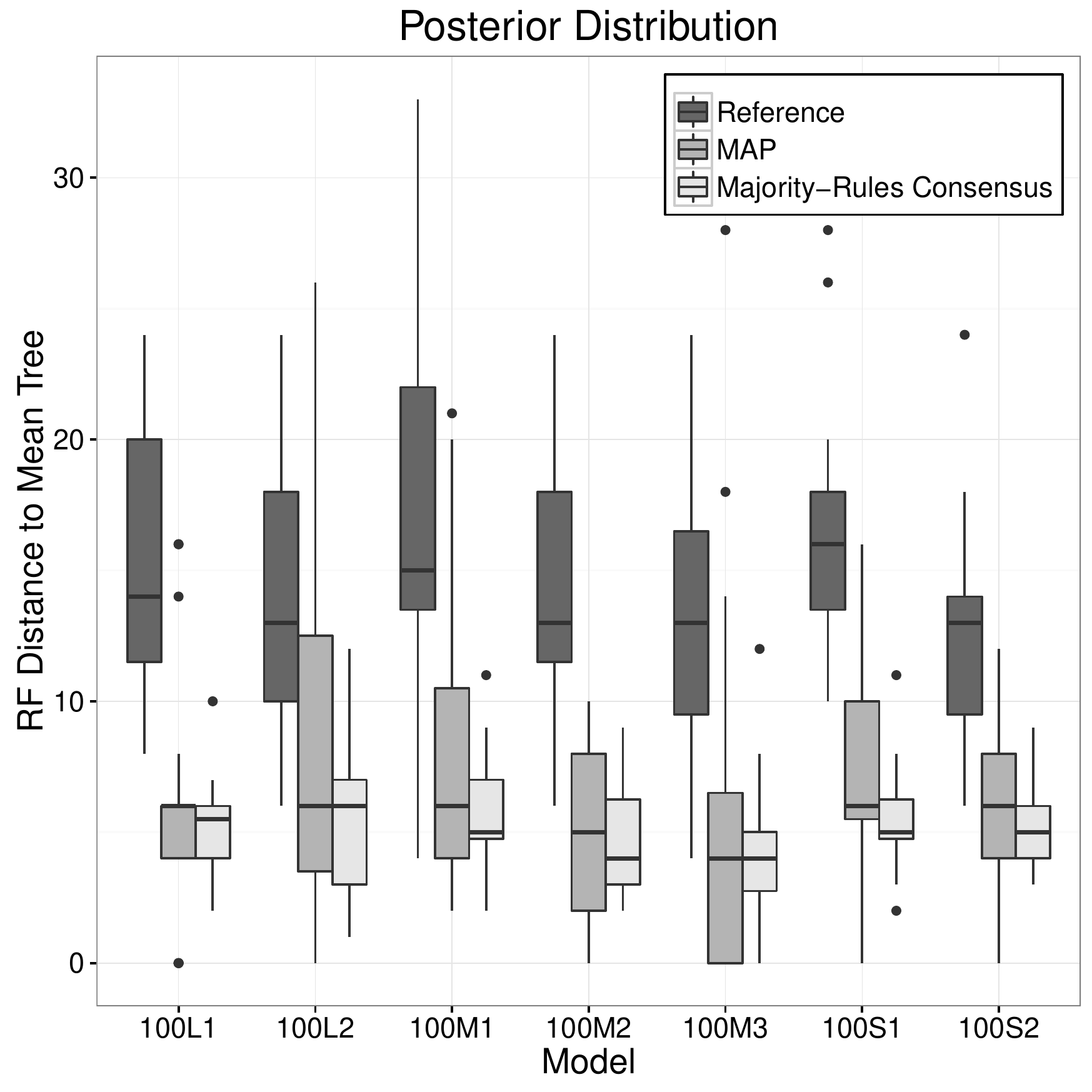} 
\includegraphics[scale=0.4]{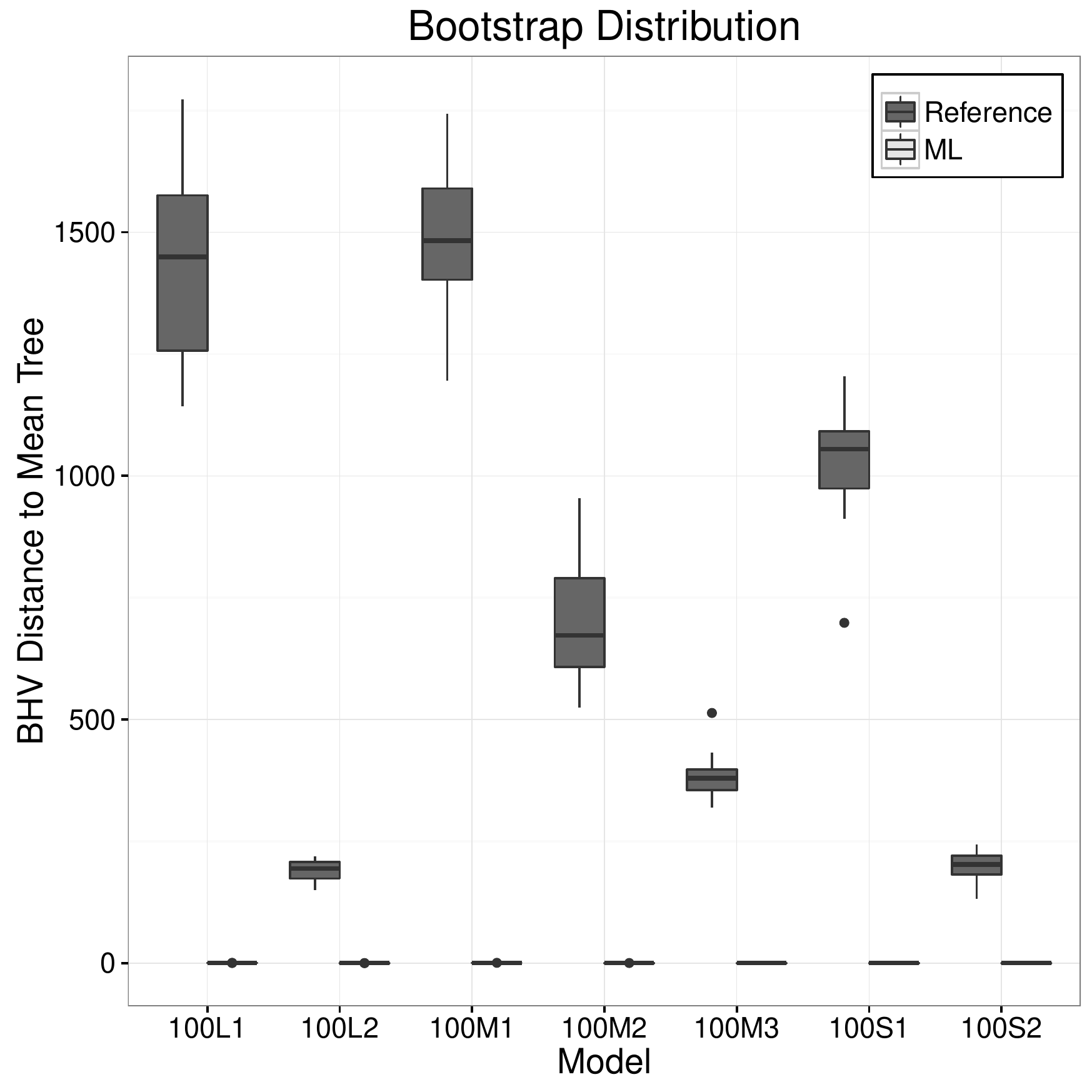} 
\includegraphics[scale=0.4]{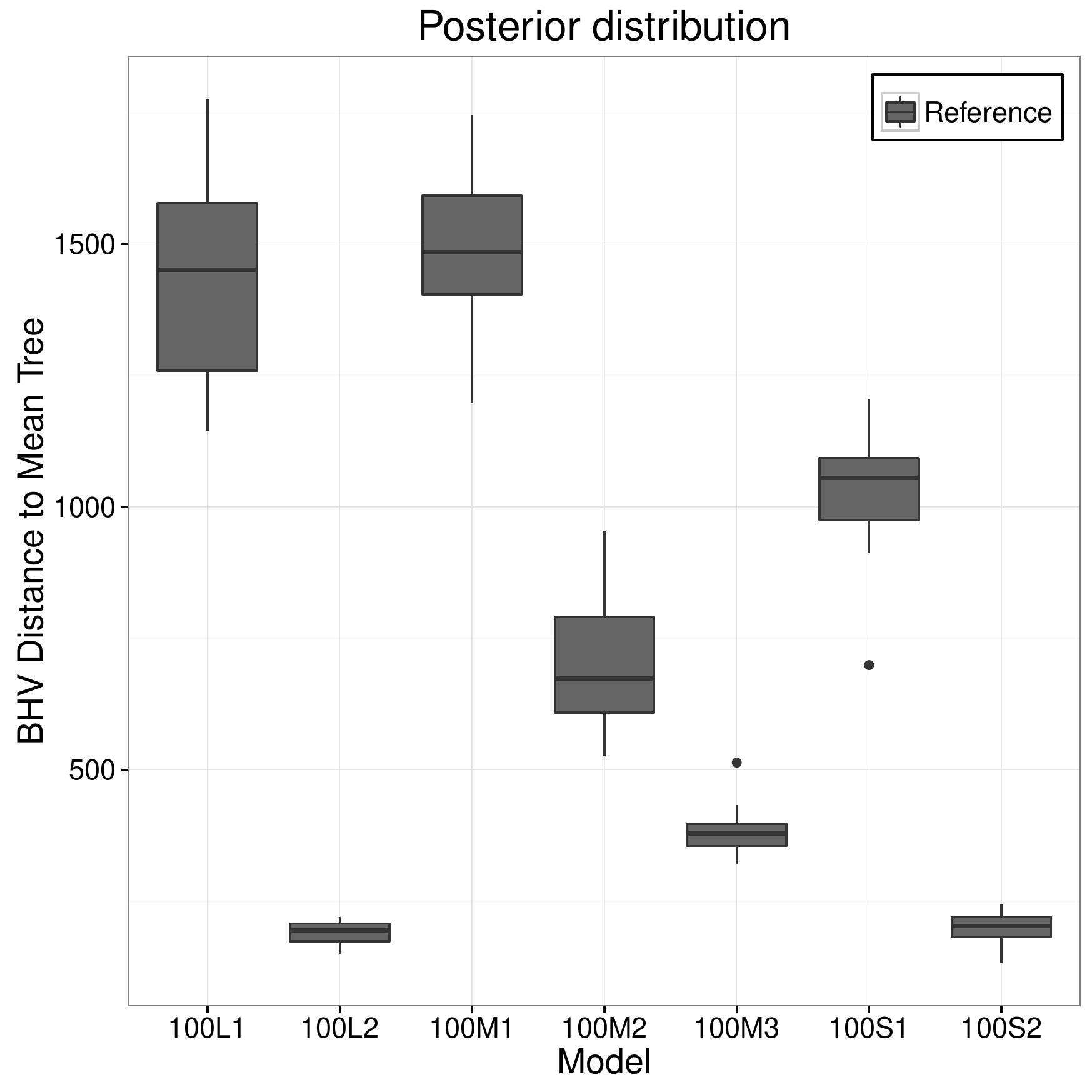}

\caption{For each sample generated from the SATE dataset, the RF and BHV distances were calculated from its mean tree to its majority-rule tree, ML or MAP tree, and reference tree. The distribution of these distances is shown using box plots. The mean tree is much closer to the ML or MAP tree and consensus tree than to the reference tree.  The variation in BHV distance between models is due to the difference in the scale of the edge lengths of the reference trees, which translates into longer edge lengths in the higher tree models.}
\centering
\label{f:sate_mean_to_all}
\end{figure}

First we compare the mean trees to the reference trees and the other summary trees, showing that they are close.  The four plots in Figures~\ref{f:mammal_mean_to_all} and \ref{f:sate_mean_to_all} use the RF and BHV distances to compare the Fr\'echet mean of each bootstrap and posterior sample to its corresponding majority-rule consensus tree, the reference tree, and either ML or MAP tree, as appropriate.  In both MAMMAL and SATE mean tree is closer to the ML or MAP and majority-rule consensus tree than to the reference tree. In the MAMMAL dataset, the distance between the mean and the other trees decreases as the sequence length increases, with those three trees almost always having the same topology for sequences of length 3000 base pairs and longer.  In the SATE dataset, the difference in the scale of the edge lengths of the reference trees results in large differences in the BHV distance between trees.

\begin{figure}[!htb]
\centering
\includegraphics[scale=0.4]{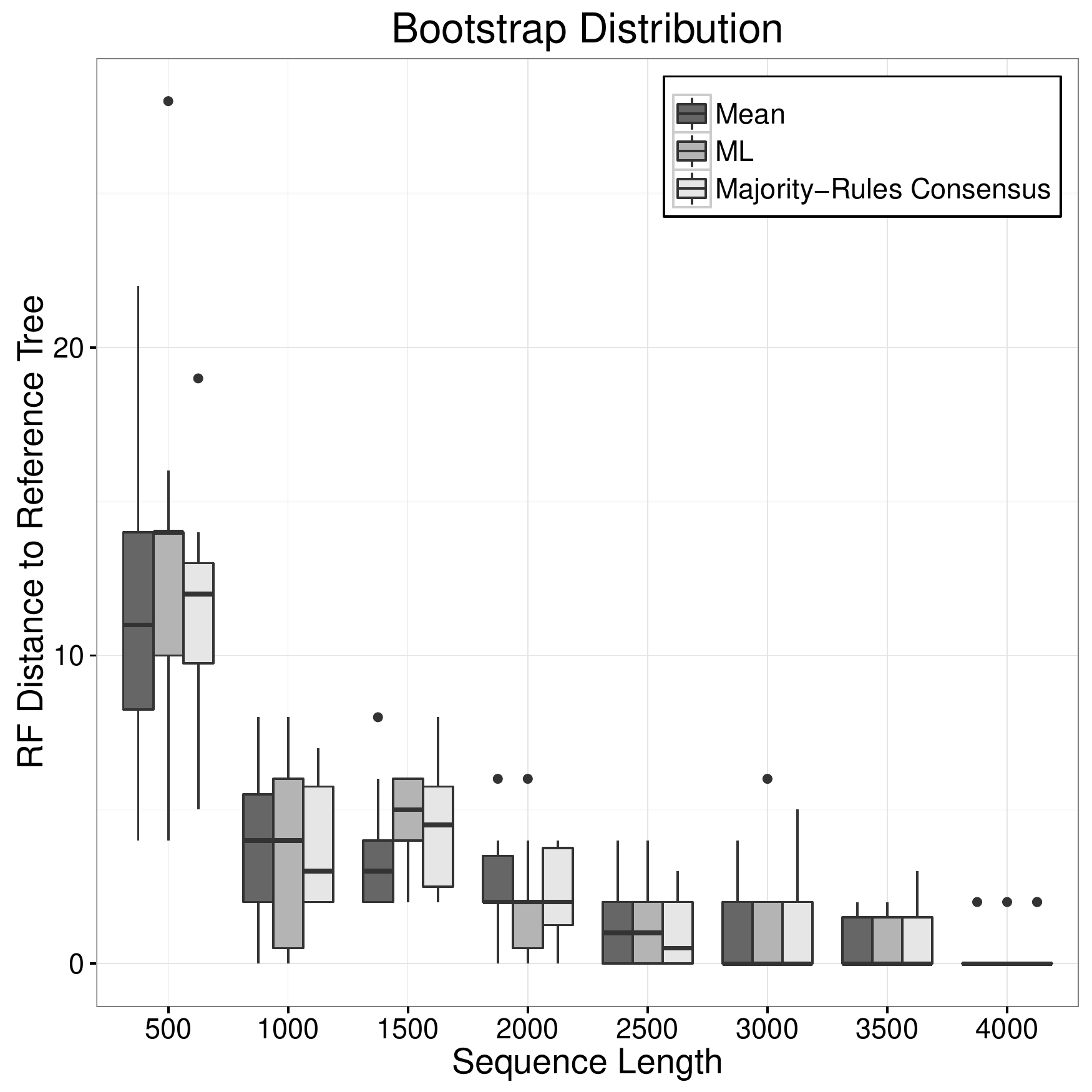}
\includegraphics[scale=0.4]{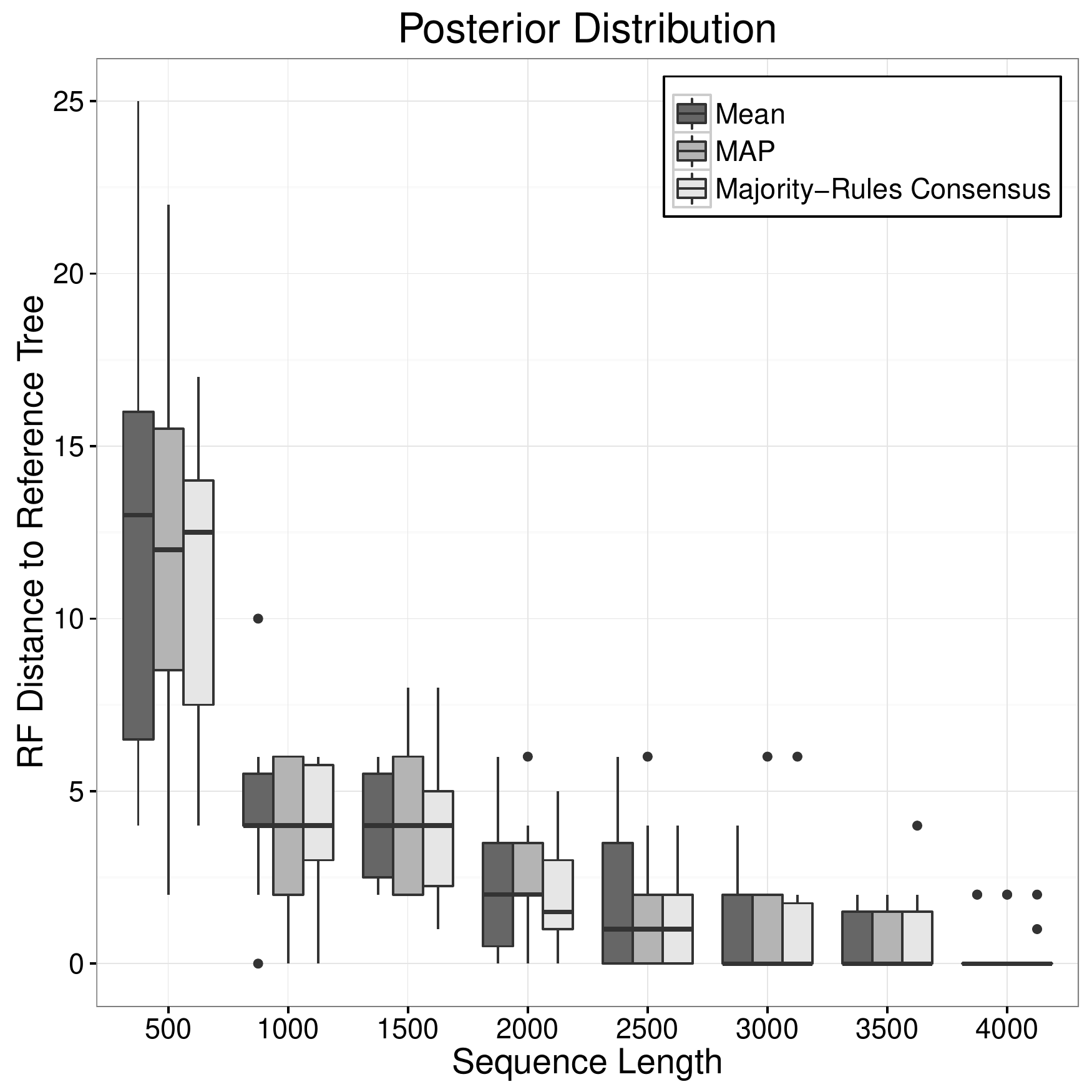}
\includegraphics[scale=0.4]{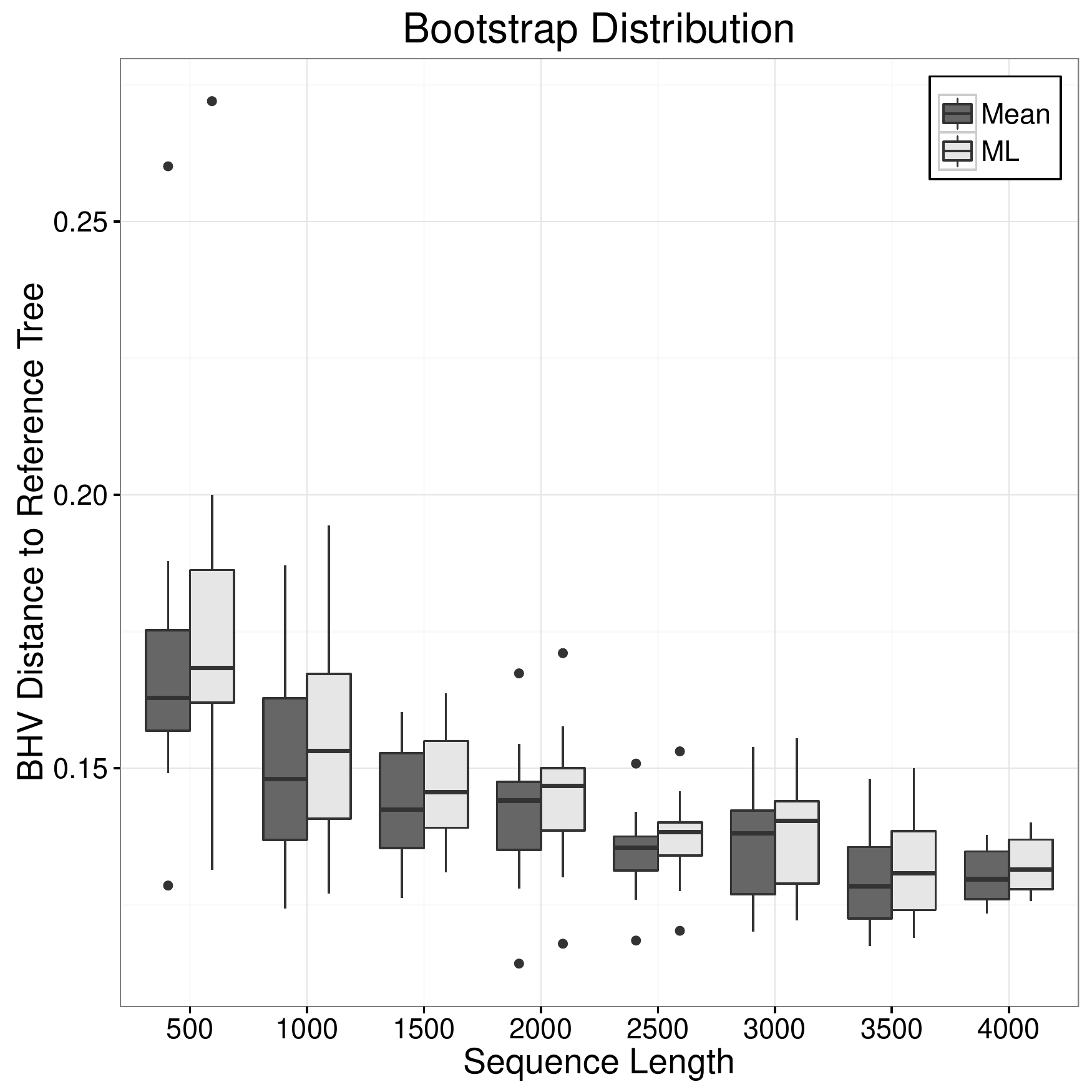} 

\caption{For each sample generated from the MAMMAL dataset, the RF and BHV distances were calculated from the reference tree to the sample's mean tree, majority-rule tree, and ML or MAP tree. The distribution of these distances is shown using box plots. All three reconstructed trees approach the reference tree as the sequence length increases, with the mean tree being slightly closer on average than the ML tree under the BHV distance measure.  The BHV distance between the reference and mean trees for the posterior distribution is shown in the lower-right graph in Figure~\ref{f:mammal_mean_to_all}.}
\centering
\label{f:mammal_orig_to_all}
\end{figure}

\begin{figure}[!htb]
\centering
\includegraphics[scale=0.4]{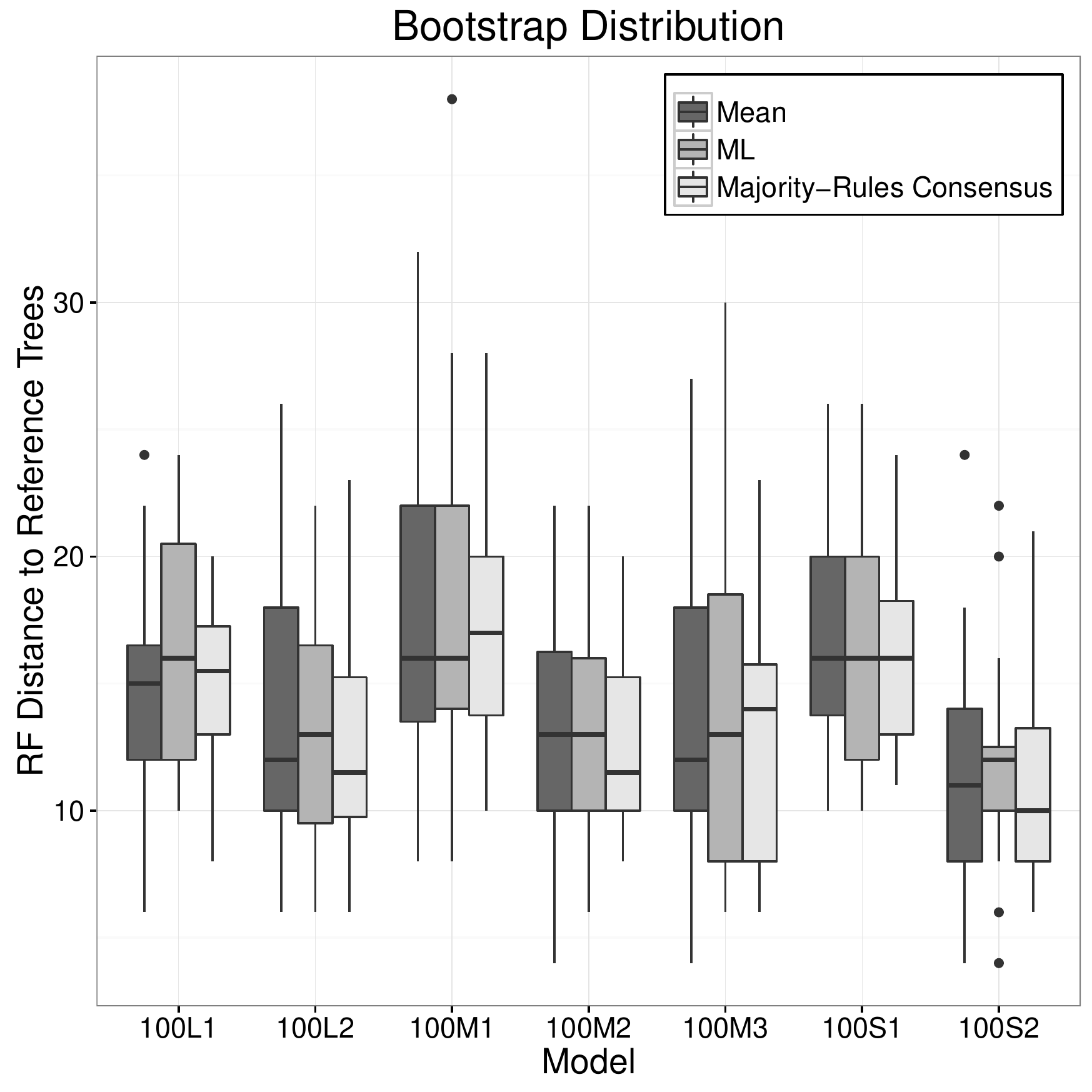}
\includegraphics[scale=0.4]{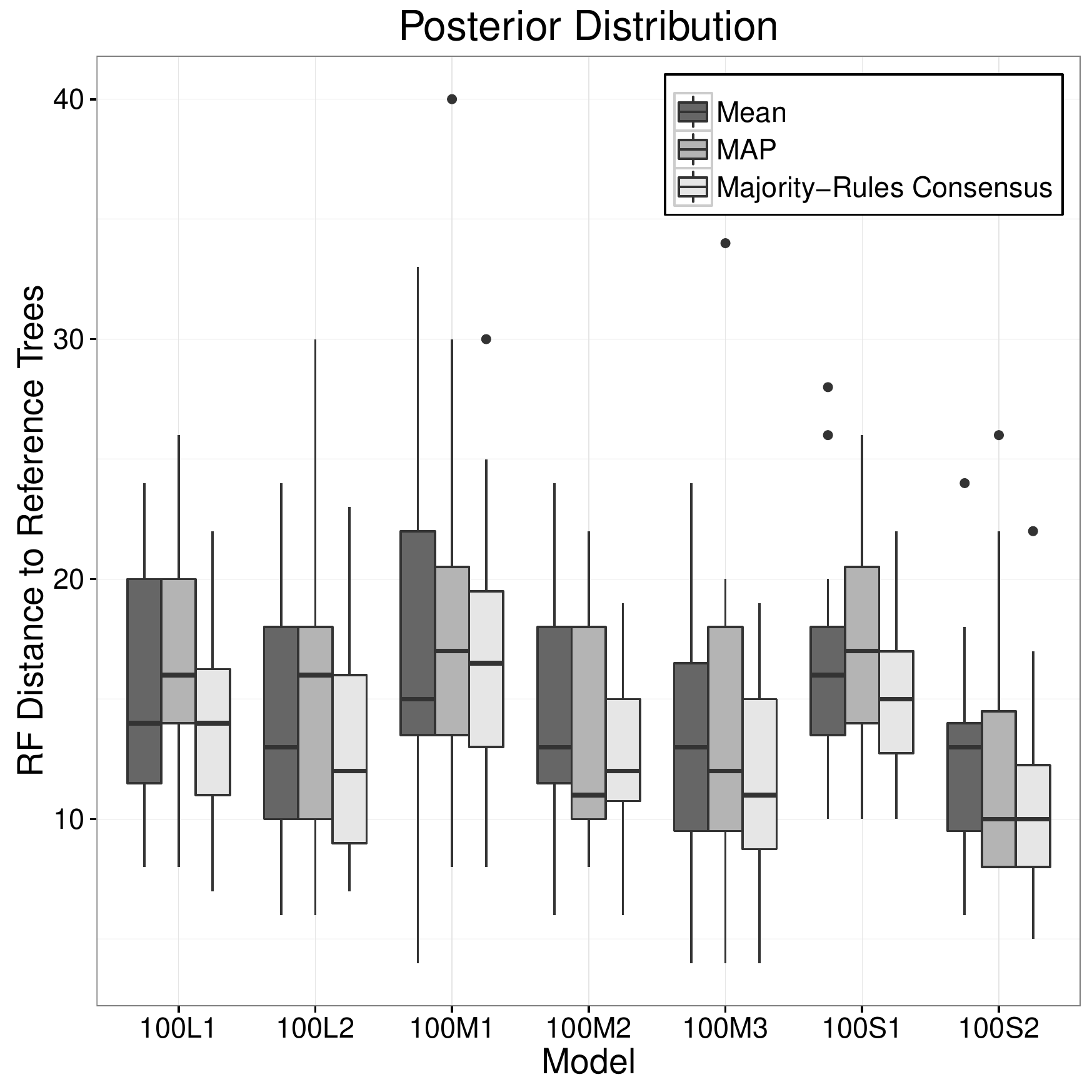}
\includegraphics[scale=0.4]{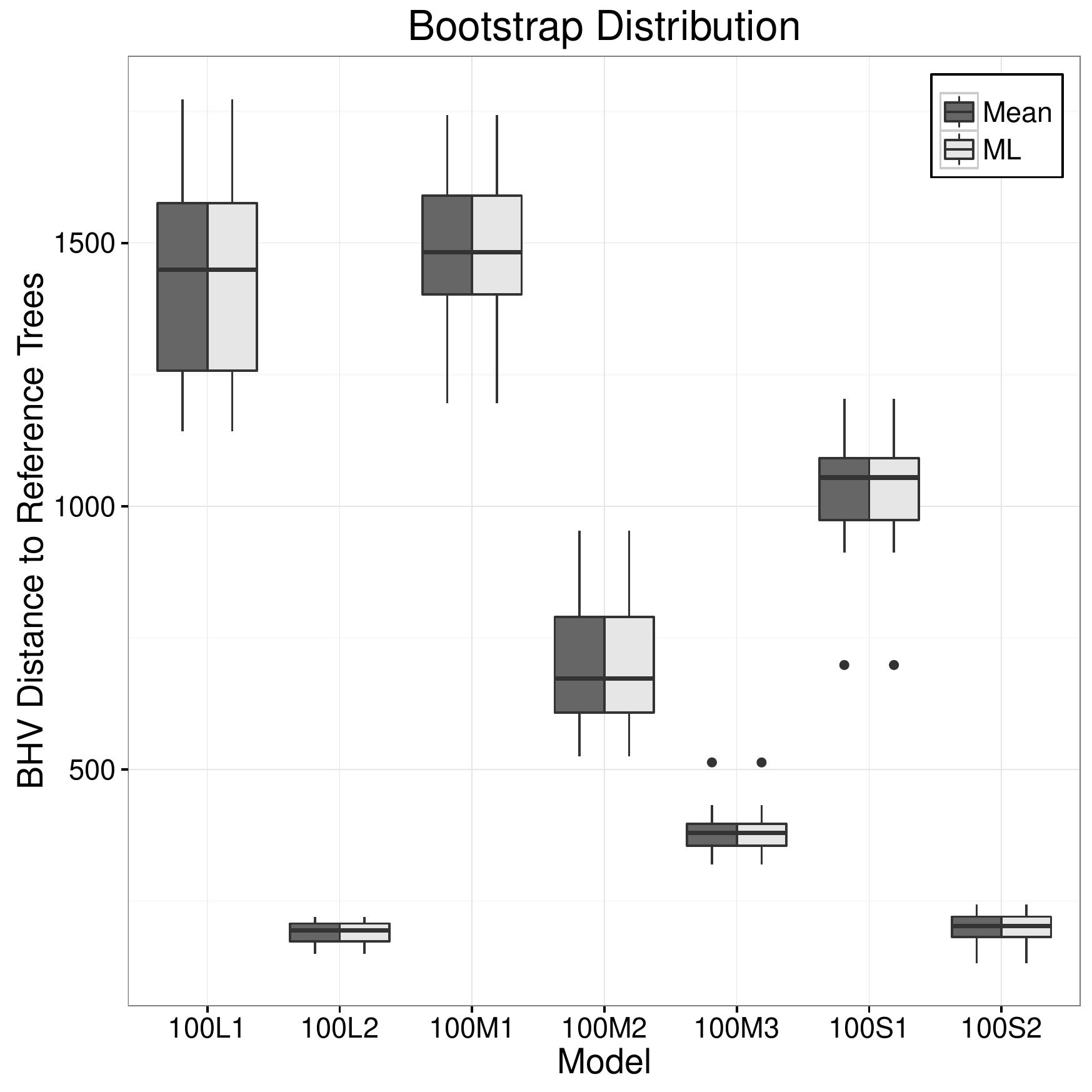}

\caption{For each sample generated from the SATE dataset, the RF and BHV distances were calculated from its reference tree to its mean tree, majority-rule tree, and ML or MAP tree. The distribution of these distances is shown using box plots. All three reconstructed trees are similar distances to the reference trees under both distances, with the variation in BHV distances due to the model variation in tree height. The BHV distance between the reference and mean trees for the posterior distribution is shown in the lower-right graph in Figure~\ref{f:sate_mean_to_all}.}
\centering
\label{f:sate_orig_to_all}
\end{figure}

Next we consider the relation of the reference trees to the reconstructed trees.  The plots in Figures~\ref{f:mammal_orig_to_all} and \ref{f:sate_orig_to_all} use the RF and BHV distance to compare the reference tree to the mean tree, majority-rule consensus tree, and ML or MAP tree of each sample.  These plots combined with the previous Figures~\ref{f:mammal_mean_to_all} and \ref{f:sate_mean_to_all} show that the reconstructed trees are closer to each other than to the reference tree.  In MAMMAL, as expected, all trees become closer to the reference tree as the sequence length increases, since this gives more information about the reference tree, improving its reconstruction.  Interestingly, under the BHV distance and for the bootstrap distribution, the mean tree is closer to the reference tree than the ML tree (p-value $< 2.2 \times 10^{-16}$ under the Wilcoxon signed rank test) for MAMMAL, but the reverse was true (p-value $< 2.2 \times 10^{-16}$ under the Wilcoxon signed rank test) for SATE.  Such strong p-values showing opposite results may indicate this behaviour is data-dependent.

\subsection{Variance}
\begin{figure}[!htb]
\centering
\includegraphics[scale=0.4]{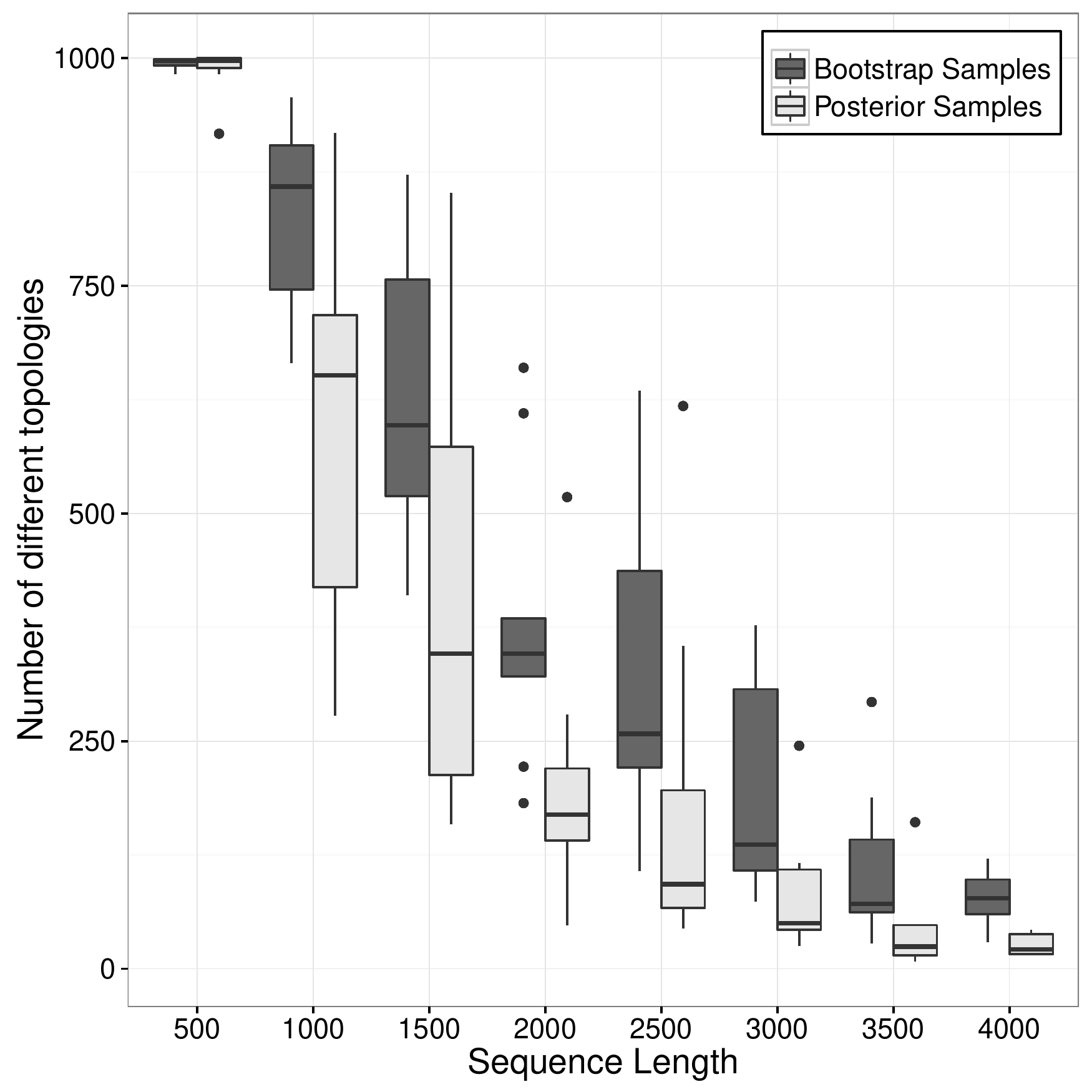}
\includegraphics[scale=0.4]{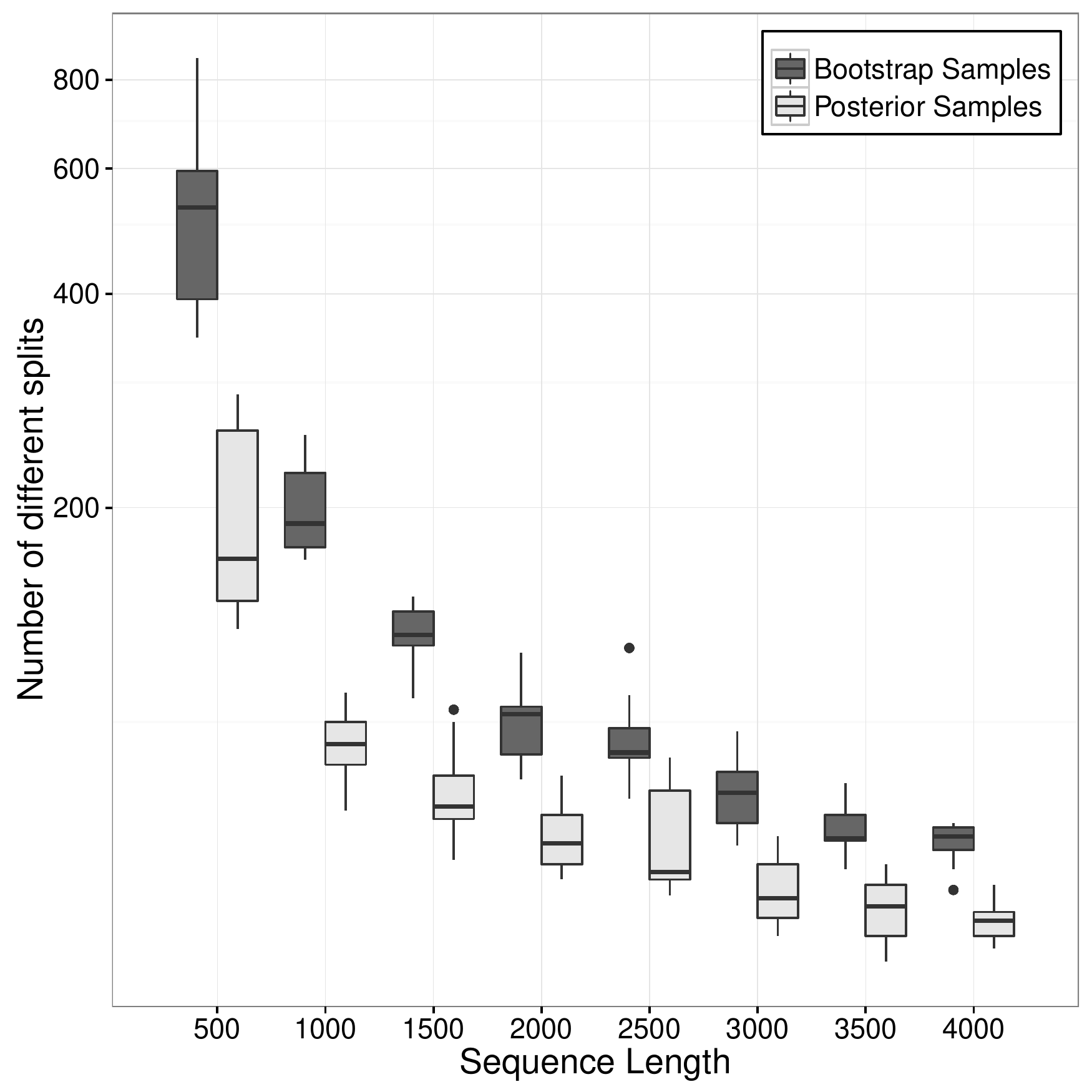} 

\includegraphics[scale=0.4]{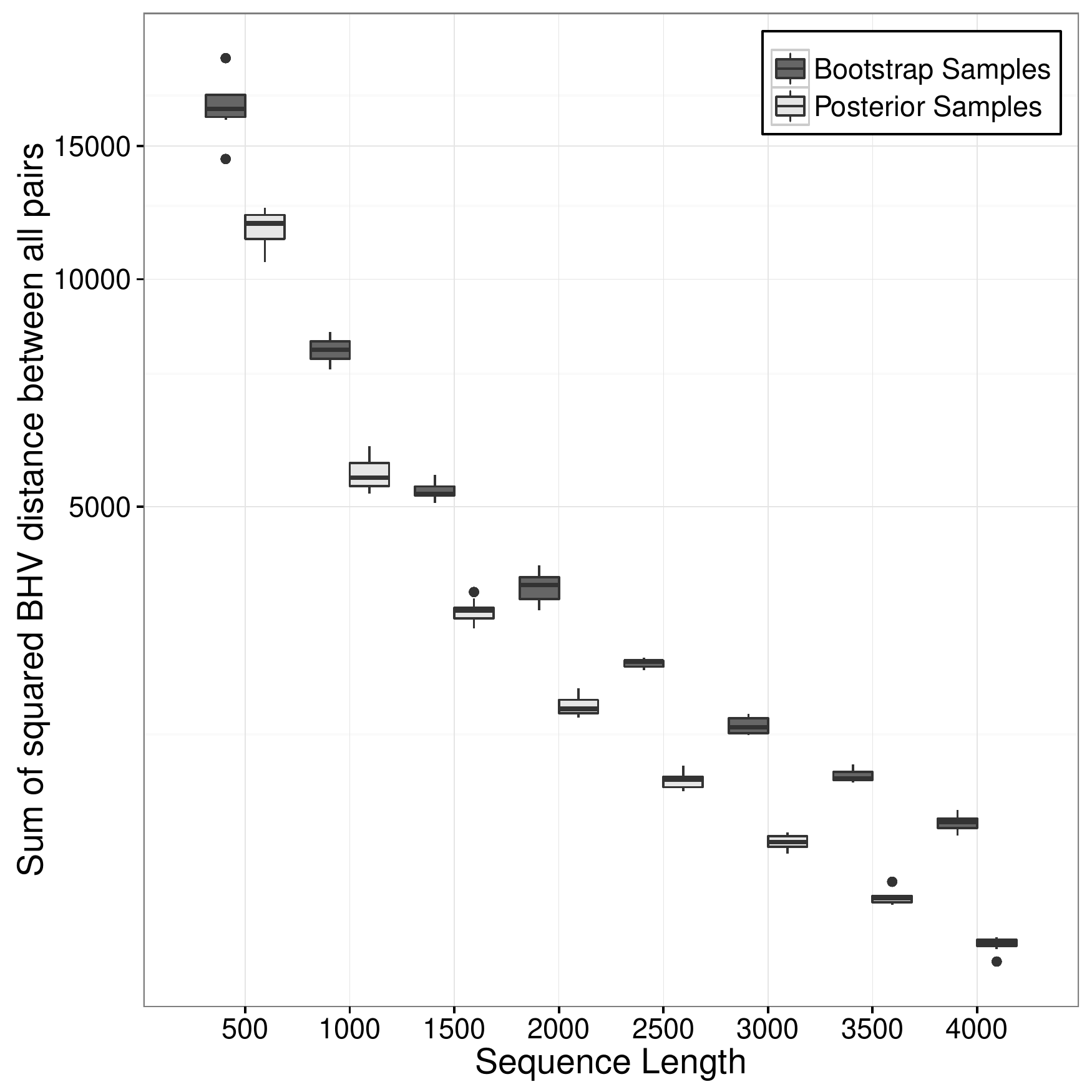}
\includegraphics[scale=0.4]{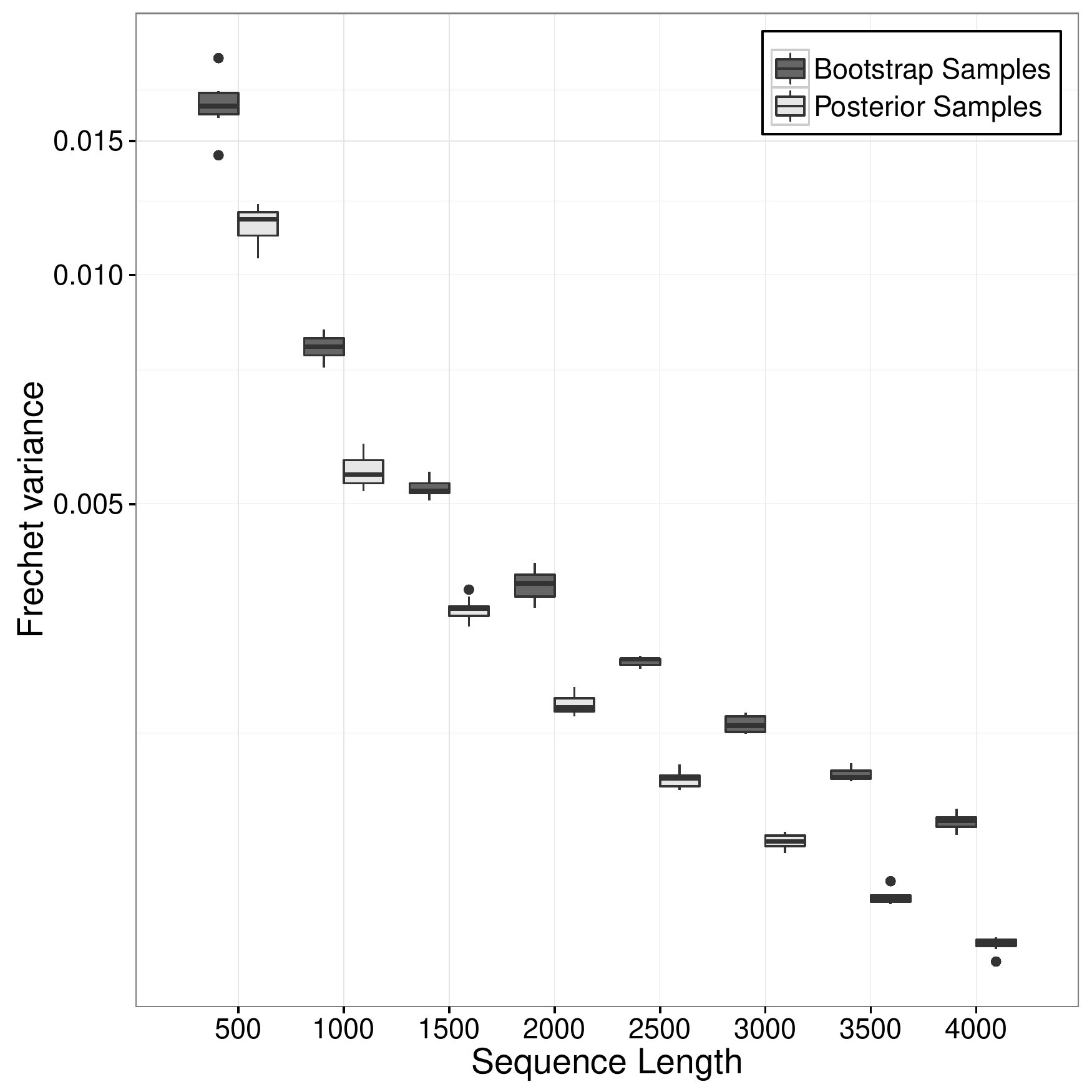} 

\caption{The variance of the MAMMAL dataset samples of the bootstrap and posterior distributions are measured using the number of different topologies, the number of different splits, the sum of the squared BHV distance between all pairs, and the Fr\'echet variance.  For all measures, the posterior samples have lower variance on average than the bootstrap samples.  The sum of the squared BHV distance between all pairs and the Fr\'echet variance are very similar, up to scaling.  Note that the y axis for the number of different splits, sum of squares, and Fr\'echet variance are in log scale.}
\centering
\label{f:mammal_all_vars}
\end{figure}

\begin{figure}[!htb]
\centering
\includegraphics[scale=0.4]{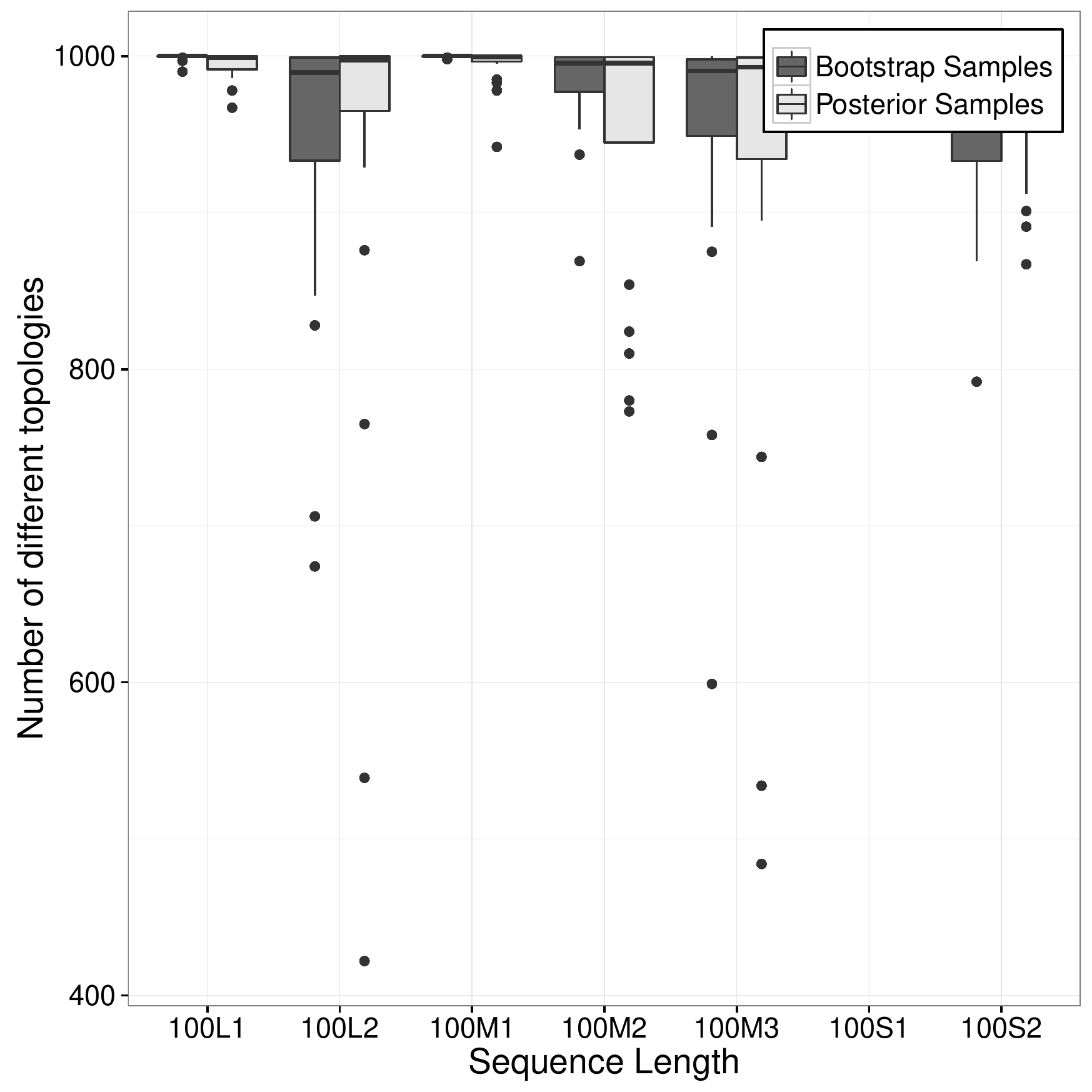}
\includegraphics[scale=0.4]{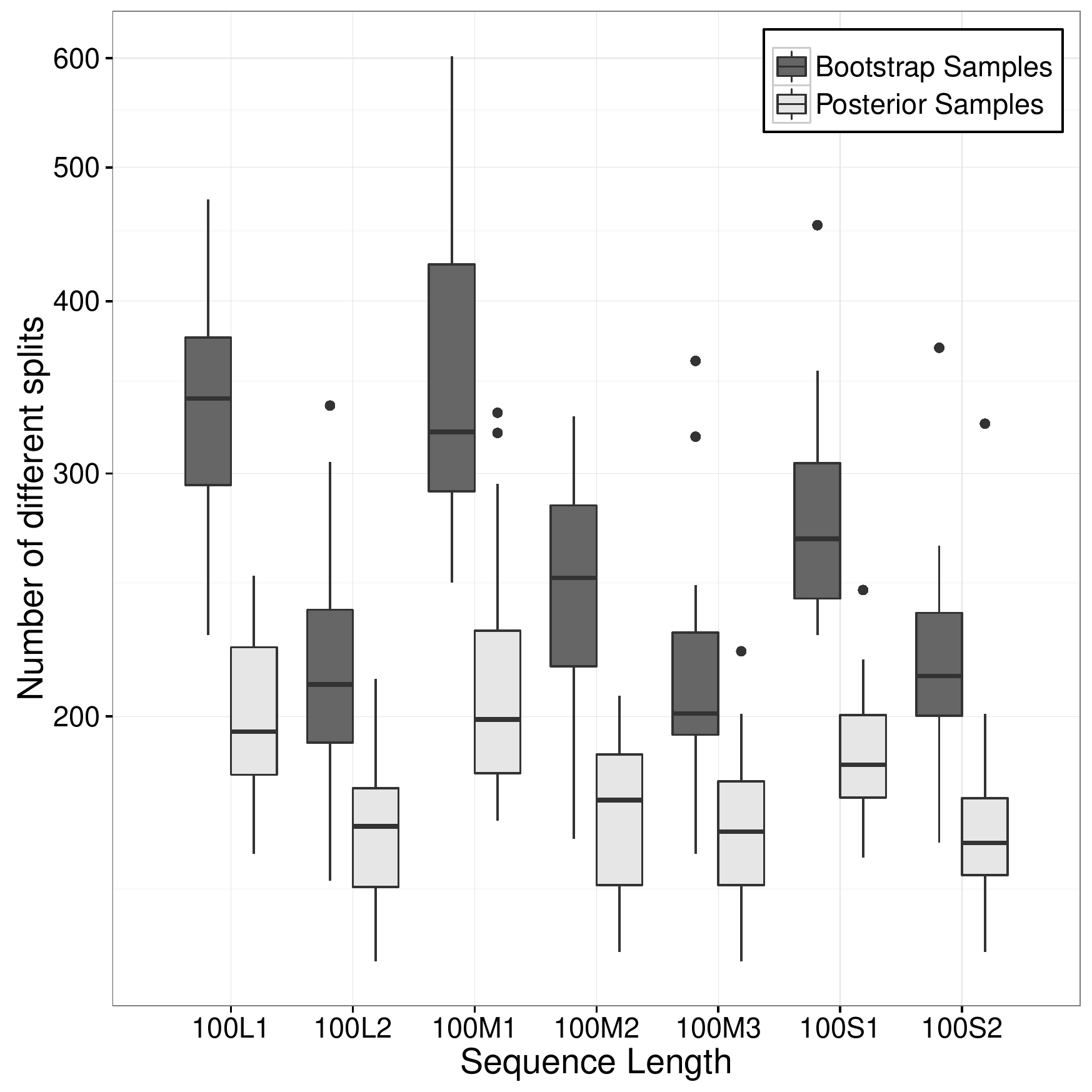} 

\includegraphics[scale=0.4]{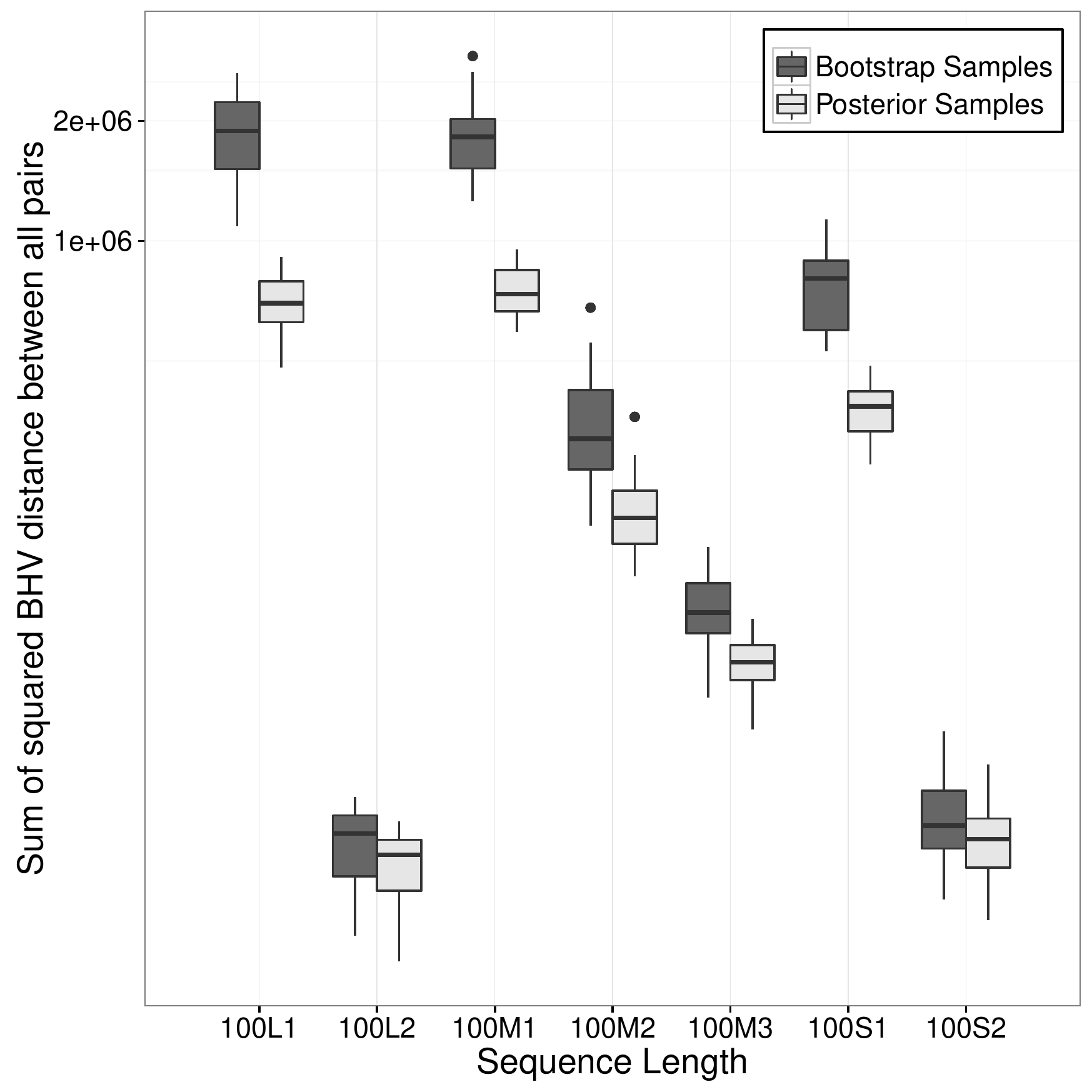}
\includegraphics[scale=0.4]{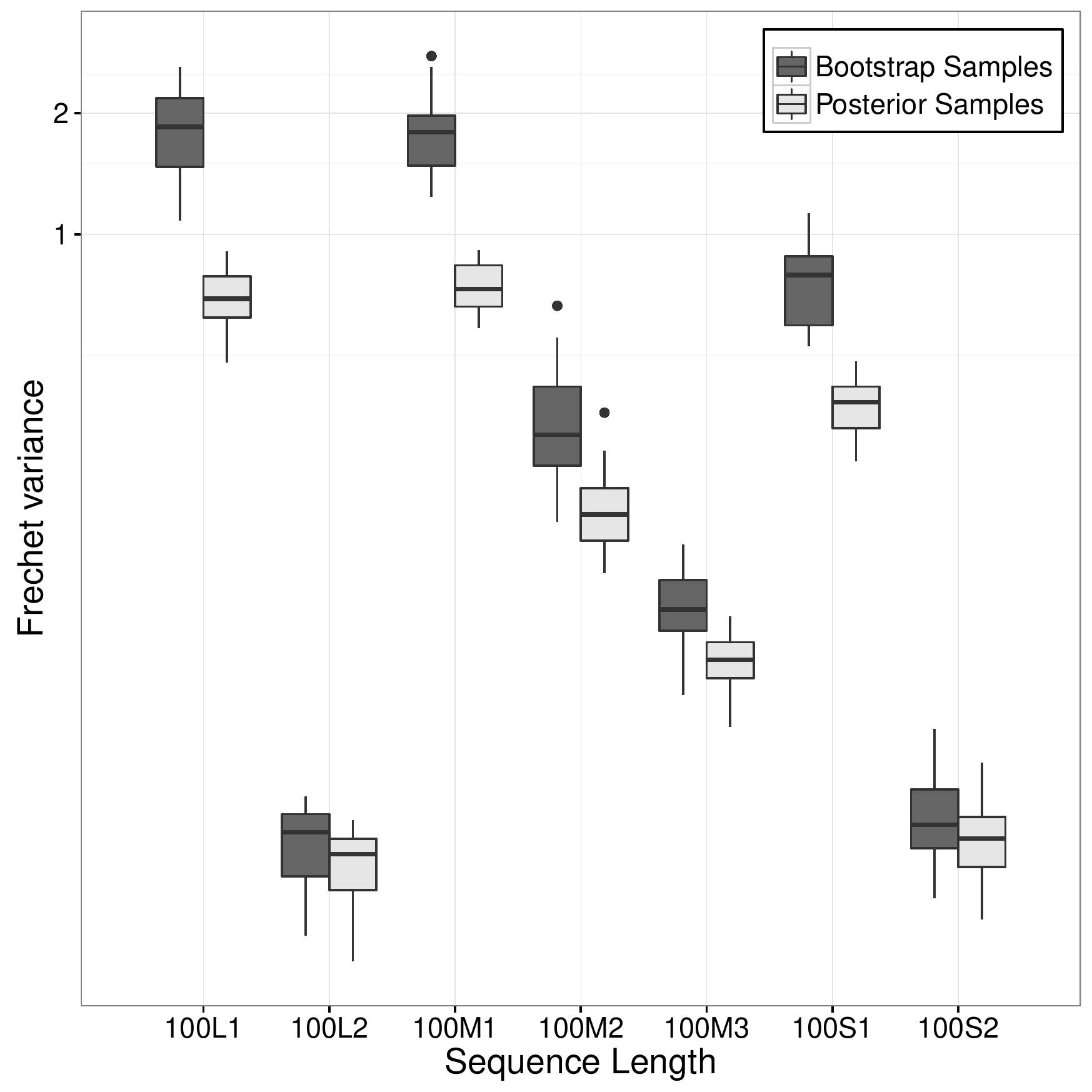} 

\caption{The variance of the SATE dataset samples of the bootstrap and posterior distributions are measured using the number of different topologies, the number of different splits, the sum of the squared BHV distance between all pairs, and the Fr\'echet variance.  For all measures, except possibly the number of different topologies, the posterior samples have lower variance on average than the bootstrap samples.  The sum of the squared BHV distance between all pairs and the Fr\'echet variance are very similar, up to scaling.  Note that the y axis for the number of different splits, sum of squares, and Fr\'echet variance are in log scale.}
\centering
\label{f:sate_all_vars}
\end{figure}

We compare the different measures of variance in a set a trees, namely the number of different topologies, the number of different splits, the Fr\'echet variance, and the sum of the 
squared BHV distance between all pairs of input trees, in Figures~\ref{f:mammal_all_vars} and \ref{f:sate_all_vars}.  In the MAMMAL dataset, as the sequence length increases, there
 is more information about the underlying tree, and so we expect RAxML and MrBayes to both do a better job at inferring this tree and be more certain about it.  This increased certainty 
 should be reflected by a decrease in variance in the bootstrap and posterior distributions, and thus samples, as the sequence length increases, which is what we see in Figure~
 \ref{f:mammal_all_vars}.  In the SATE dataset, the variance decreased as the amount the edge lengths are scaled by decreased.  Only model pairs 100L1 and 100M1, and 100L2 and 
 100S2 have different gap lengths, but had edge lengths scaled by the amount.  However, this change in gap length does not appear to have affected the variance. For both datasets, 
 for all measures, expect possibly the number of different topologies for SATE, the posterior samples have a lower variance than the bootstrap samples. This lower variance matches 
 previous observations that the posterior probabilities are higher than bootstrap frequencies for well-supported clades 
 \citep{erixon2003reliability,douady2003comparison,huelsenbeck2004frequentist}, since trees in a lower variance sample are not as spread out, and thus have fewer different splits, 
 than trees in a higher variance sample.  In general, we find that there is more uncertainty in the variance estimate for the other metrics than the Fr\'echet variance.

The measurements of the sum of the squared BHV distance between all pairs of sample trees and the Fr\'echet variance are almost identical, up to scaling.  This similar behaviour is expected as these two measures are equivalent in Euclidean space, which is why the sum of squared BHV distance between all pairs was first suggested as a measure of variance.  For large input sets, it is faster to compute the Fr\'echet variance than the BHV distance between all pairs.  See the Discussion, below, for more details.

 \bigskip
\subsection{Unresolved Trees and Tree Length Comparison}
\begin{figure}
\centering
\includegraphics[scale=0.5]{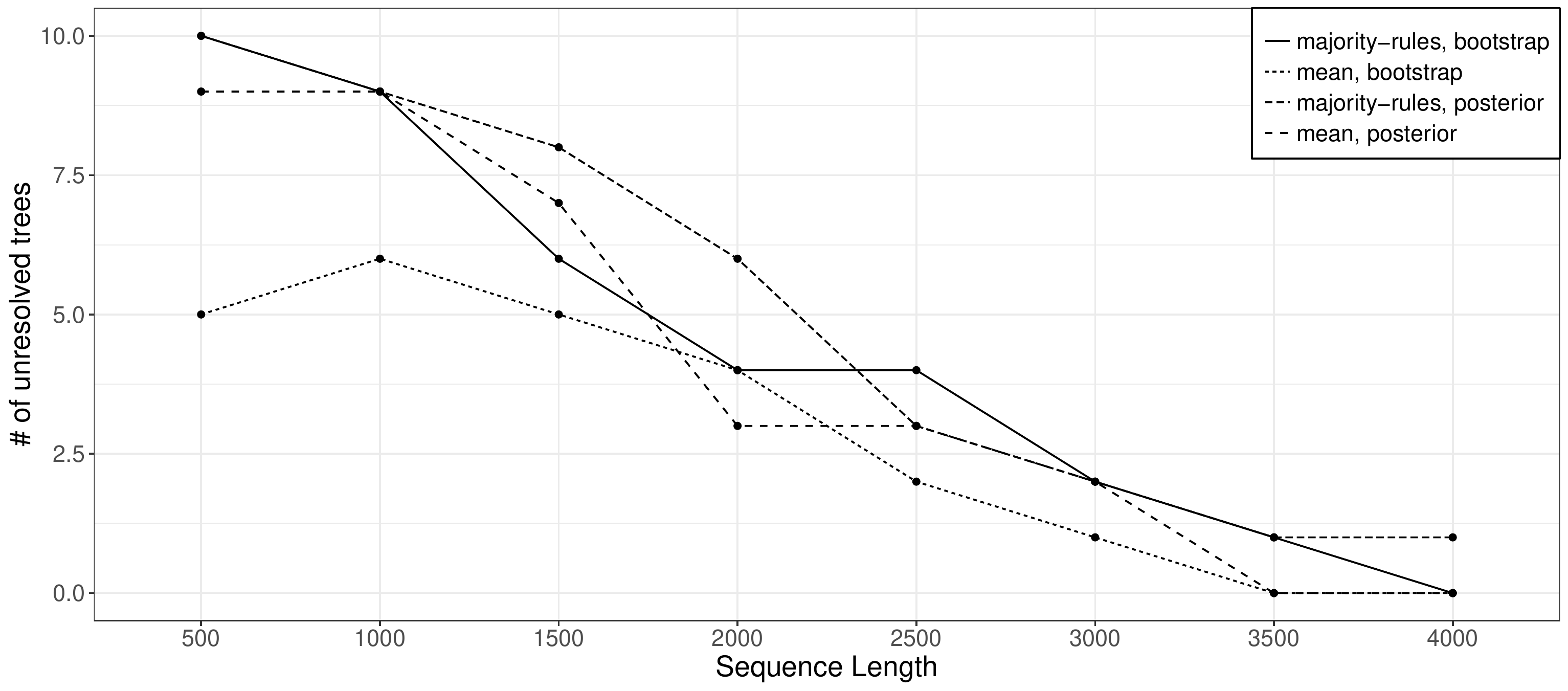}
\caption{In the MAMMAL dataset, there were the same number or fewer unresolved mean tree than majority-rules trees for the bootstrap and posterior samples for all sequence lengths.  The number of unresolved mean trees decreases as the sequence length increases, reaching 0 for 4000 base pairs.}
\centering
\label{f:unresolve}
\end{figure}

We also looked at the average tree lengths and how many mean and majority-rules trees were unresolved to determine the effect of mean stickiness.  Figure~\ref{f:unresolve} shows the number of unresolved mean and majority-rules trees for the bootstrap and posterior samples for all sequence lengths in MAMMAL.  The number of unresolved mean and majority-rules trees decreases as the sequence lengths increase, with all mean trees for the 4000 base pair sequence length being binary.  There are always the same or fewer unresolved mean trees than unresolved majority-rules trees for each sequence length.  However, even when mean trees are fully resolved, the stickiness property could still cause them to be closer to the origin of the sample than expected. In terms of measurable effects, being closer to the origin translates into the mean trees being shorter (in terms of total edge length) on average than the trees from which they were computed.  For each of the 80 (10 replicates per sequence length) bootstrap and posterior samples, the length of the mean tree was strictly less than the average length of the corresponding sample trees.
While the mean trees had a shorter length than might be expected, this does not noticeably affect the quality of the summary, as averaged over all sequence lengths, the bootstrap and posterior mean trees were only 99.48\% and 99.50\%, respectively, of the average lengths of their respective samples. 

We also compared the lengths of the ML trees of the bootstrap samples with their corresponding mean trees for the MAMMAL dataset.  In all 80 samples the length of the ML tree was greater than the length of the corresponding mean tree.  With $p$-value 0.005861, the length of the ML tree is greater than the average length of the trees in the corresponding bootstrap sample.

Interestingly, in the MAMMAL dataset, the average length of the trees in each posterior sample was always less than the average length of the trees in the corresponding bootstrap sample generated from the same set of sequences. This difference suggests the bootstrap samples are more spread out, which could also reflect posterior probabilities being higher for well-support clades than bootstrap probabilities.  Unsurprisingly, the mean tree of the posterior distribution sample always had length less than the mean tree of the corresponding bootstrap sample. 

\subsection{Mean Hypothesis Test}

For the first repetition of the 4000 base pair sequences, we performed a two-sample hypothesis test to compare the samples of the bootstrap and posterior distributions using the means.  The null hypothesis stated that the two means of bootstrap and posterior distributions were the same.  We tested this hypothesis using an approximate permutation test.  In all cases, the distance between the means of a random partition of the two samples was strictly less than the distance between the means of the two samples themselves.  Thus we reject the null hypothesis with an estimated p-value of 0.002, with a 95\% confidence interval of $[0,0.006]$.  Therefore we can assume the two distributions do not have the same mean, and thus are not the same.  This result was expected due to previous work showing the bootstrap and posterior probabilities are different \citep{erixon2003reliability,douady2003comparison,huelsenbeck2004frequentist}.


\bigskip
\section{Discussion}
\label{s:discussion}





Our experiments demonstrate that both the Fr\'echet mean and variance behave in the expected way on biological data. We have shown that the mean of samples of bootstrap and posterior distributions is comparable in accuracy to the ML and MAP trees, respectively, as well as the majority-rule consensus topology. Furthermore, the mean tree is more likely to be binary than the majority-rules consensus topologies. In the MAMMAL dataset, the original reference tree is closer to the mean than to the ML tree, however the reverse is true in the SATE dataset.  This leads to the open question of whether we can characterize when the Fr\'echet mean performs better than the ML tree as a summary for bootstrap samples.  While this possible gain may not be enough to warrant the cost of computing the mean, these results conclusively demonstrate that the mean is a valid summary method for a sample of trees.  We believe the value of the mean comes from its sound mathematical backing, which enables more sophisticated statistical tests, like mean hypothesis testing.

From the results of the variance experiments, it is clear that the Fr\'echet variance is a faster and more precise measure of variance than existing alternatives. The variability in the number of topologies in the tree samples with the same sequence length is very high, in comparison to the other variance measures. Although the variability decreases when measuring the number of splits in the tree sample, it is still higher than that of the sum of squared pairwise distances and the Fr\'echet variance.  {From the SATE dataset variances, we see that tree length can be a key, and possibly overwhelming, factor in the value of the variance.  However, the MAMMAL dataset experiments show that when tree samples are comparable in length, the variance provides a means of differentiating between sample variability.  Thus, the variance will be most useful in comparing samples of similar average tree length.  For example, when comparing the bootstrap or posterior samples for a number of genes on the same taxa, high variance relative to average tree length of a sample could indicate an outlier. 

The sum of squared pairwise distances and the Fr\'echet variance have very similar profiles, up to scaling, justifying the use of the sum of squared pairwise distances as a measure of variance in the literature.  However for large sample sizes, the Fr\'echet variance is faster to compute.  To compute the sum of squared geodesic distances for $r$ trees, one must compute $r(r-1)/2$ geodesic distances.  In contrast, computing the Fr\'echet variance involves calculating the geodesic distance once per iteration of the algorithm.  The number of iterations required depends on the desired precision of the mean.  However, we have often obtained good results with 10,000-15,000 iterations, suggesting that when we are computing variances of more than approximately 200 trees, the Fr\'echet variance calculation will be faster than the sum-of-squared-distance calculation.
 
Our experiments on the tree lengths show that while the mean trees are shorter than the average length of the sample trees, it is by less than 1\% on average. This demonstrates that stickiness leads to shorter mean trees, but that the difference in lengths is not likely to be significant and can be ignored.  More importantly, we showed that the mean tree is less sticky than the majority-rules topology, because it has the same or fewer unresolved edges.  The Euclidean median is sticky, while the Euclidean mean is not, so since the majority-rules topology is the median under the RF distance, it is likely more sticky than the mean tree.  Thus, using the mean, instead of majority-rules topology, to summarize a set of trees gives more resolution, albeit at higher computation cost.  Note that the iterative algorithm for approximating the mean works in such a way that it almost always produces binary trees, so any edges of length less than the approximation amount should be trimmed.  

One might think that we could use the mean to estimate the species tree for a set of gene trees \cite[Example 5.5]{MillerOwenProvan2015}.  Even if we just restrict ourselves to the coalescent model for explaining gene tree diversity, this estimation is still problematic in two regards.  Under the coalescent model, the pendant edges of the gene trees must be at least as long as than the pendant edges of the species tree, but will usually be longer.  Since each pendant edge of the mean tree is computed by averaging the length of that edge in all input trees, the pendant edges of the mean of the gene trees will be at least as long, and usually longer, than the pendant edges of the true species tree.  While it is possible that the topology of the mean tree might still match that of the species tree, as conjectured in \cite[Example 5.5]{MillerOwenProvan2015}, our preliminary experiments suggest that stickiness greatly limits the amount of information in the mean tree, compared to other methods.  In preliminary experiments, the mean tree of a set of simulated gene trees became essentially unresolved while the gene trees were still far away from the anomalous zone \citep{DegnanRosenbergAnomalous}.  That is, the most common gene tree topology was still the same as the species tree topology.  By essentially unresolved, we mean that the approximated mean tree was within a small distance of some unresolved.  It is possible that an exact algorithm for computing the mean would give a binary mean tree topology even in these cases, but with an iterative algorithm, we can not be certain the mean is not unresolved, nor even that it would be informative.

\bigskip
\section{Conclusion}

We have shown that the Fr\'echet mean and variance behave in an expected way on biological data.  We have further shown that the mean is more likely to be resolved than the majority-rules tree, and that the variance is a stable and reliable measure of the amount of variability in a sample of trees.  This validation opens the door to new applications of these quantities.

One possible application of the variance is to determine when MrBayes or other Markov Chain Monte Carlo (MCMC) algorithms converge.  Preliminary experiments on the MrBayes runs for the data in this paper show that comparing the variance of sliding windows of the sampled trees can identify the burn-in, but that the variance of the trees in the sliding window remains roughly the same after this period. However, there may be other datasets where the variance will continue to decrease after the burn-in, indicating a lack of convergence.  The iterative algorithm for computing the mean is easily adapted into an online algorithm:  at each iteration, instead of choosing a  tree at random from the input set of trees, use the next tree generated by the MCMC chain.  Unfortunately, it is not clear how to take advantage of this online algorithm in computing the variance, which requires computing the BHV distance from the current mean approximation to all sample trees seen so far.  However, algorithms for computing geodesics dynamically \citep{SkwererProvan_dynamicgeos} might help.

\section{Acknowledgements}
We would like to thank the anonymous reviewers for their comments and suggestions that greatly improved the quality of the paper. We also thank Katherine St. John, Tandy Warnow, Derrick Zwickl, Lior Pachter, Tom Nye, and Aasa Feragen for helpful discussions.  MO acknowledges the support of the Fields Institute. This work was partially supported by a grant from the Simons Foundation (\#355824, Megan Owen).  The work of DGB is supported by an NSERC Discovery Grant.


\bigskip\bigskip


\bibliographystyle{sysbio}
\bibliography{meanvar}







\end{document}